%% file: main.tex
\documentclass[sigplan,nonacm]{acmart}

\input{sage_macros}

\usepackage{array}
\usepackage{booktabs}
\usepackage{tabularx}
\usepackage{libertine}
\usepackage[libertine]{newtxmath}
\usepackage{xcolor}
\usepackage{pgfplots}
\pgfplotsset{compat=1.18}
\usepackage{graphicx,tikz}
\usetikzlibrary{arrows.meta,calc,decorations.pathreplacing}

\begin{document}

\title[SAGE]{\sage: Semantic-Aware Geographic Error Recovery for AI Data Movement}

% \author{Anonymous Author(s)}

\author{Patrick S. Y. Hung}
\affiliation{%
  \institution{City University of Hong Kong}
  \country{Hong Kong}
}
\email{psyhung@cityu.edu.hk}

\author{Zitong Wang}
\affiliation{%
  \institution{City University of Hong Kong}
  \country{Hong Kong}
}
\email{zitowang7-c@my.cityu.edu.hk}

\author{Zekai Zhang}
\affiliation{%
  \institution{City University of Hong Kong}
  \country{Hong Kong}
}
\email{zekazhang2-c@my.cityu.edu.hk}

\author{Yu Hin Chan}
\affiliation{%
  \institution{City University of Hong Kong}
  \country{Hong Kong}
}
\email{yhchan96@cityu.edu.hk}

\author{Shengzhe Lyu}
\affiliation{%
  \institution{City University of Hong Kong}
  \country{Hong Kong}
}
\email{shengzhe.lyu@my.cityu.edu.hk}

\author{Ray C.C. Cheung}
\affiliation{%
  \institution{City University of Hong Kong}
  \country{Hong Kong}
}
\email{r.cheung@cityu.edu.hk}

\renewcommand{\shortauthors}{Hung et al.}

\input{sections/abstract}

\maketitle

\input{sections/introduction}
\input{sections/contract}
\input{sections/architecture}

\input{sections/implementation}

\input{sections/methodology}
\input{sections/evaluation}

\input{sections/related_work}
\input{sections/conclusion}

% Start the uncounted reference section on a fresh page so the
% 11-page main-paper boundary is visually unambiguous.
\clearpage
\bibliographystyle{ACM-Reference-Format}
\bibliography{references}

% ASPLOS permits anonymized appendices in the submission PDF, but the
% main paper must remain self-contained.  Start them after references.
\appendix
\input{appendix/appendix_a}
\input{appendix/appendix_b}

\end{document}

%% file: sections/abstract.tex
\begin{abstract}
AI interconnects typically protect and replay packets uniformly, yet numerical bit faults differ sharply in consequence: a low-order mantissa flip may resemble quantization noise, while a high-significance exponent flip can produce a catastrophic outlier or non-finite value.
We present \sage, a semantic-aware geographic error-recovery architecture that decouples whether a detected fault merits replay from where replay restarts.
For BF16-like data, a workload-calibrated contract separates catastrophic Class-\clH{} faults from bounded Class-\clM{} and precision Class-\clL{} damage.
It first applies a Class-\clH{} silent-delivery constraint, then ranks admissible policies by quality-normalized terminal latency, $\Psi_{\rm del}$.
Independently, a source-local region table adapts checkpoint intervals to fault geography, shortening recovery segments in noisy regions.
Detected Class-\clH{} failures may trigger protected negative acknowledgments and full-flit replay; Class-\clM{} and Class-\clL{} outcomes do not trigger default network replay.

We implement \sage's endpoint and replay protocol in gem5 Garnet and synthesize its fully pipelined checker in ASAP7.
At a stable synthetic operating point, a ten-seed contention-faithful direct-Garnet campaign shows that \sage reduces $\Psi_{\rm del}$ by 30.1\% relative to fixed 34-hop recovery, combining 28.0\% lower mean latency with improved delivered semantic quality.
Under higher-BER synthetic stress at the same offered load, \sage maintains bounded queues while the fixed baseline accumulates backlog.
Application-derived DeiT-S communication traces also show lower mean and p99.5 latency at the evaluated nonzero BERs.
Within the qualified operating envelope, CRC32 decoder trials yield a simultaneous 95\% per-original Class-\clH{} silent-delivery upper bound of $3.18\times10^{-7}$.
\end{abstract}

%% file: sections/introduction.tex
\section{Introduction}
\label{sec:introduction}

AI tensors traverse networks-on-chip (NoCs), chip-to-chip links, wafer-scale fabrics~\cite{rashidi2025fred}, and high-bandwidth memory interfaces exposed to physical faults~\cite{FlynnHung2005RoadAhead,benini2002noc}.
Yet bit faults differ sharply in numerical consequence: a low-order mantissa fault may resemble precision noise, whereas a high-significance exponent fault can create an outlier or non-finite value that dominates a reduction or corrupts model state~\cite{li2017errorpropagation,reagen2018ares,rakin2019bitflip}.

In our DenseNet-121/ESC-50 fault-injection experiment, exponent-MSB faults cause catastrophic training collapse: accuracy falls to 2.0\% (chance) and remains there at every subsequent measured checkpoint, compared with 83.3\% for the clean control (Section~\ref{sec:evaluation-semantic}).
Recovery should therefore account for numerical consequence, not just whether a bit was corrupted.

\begin{figure*}[t]
    \centering
    \includegraphics[width=\textwidth]{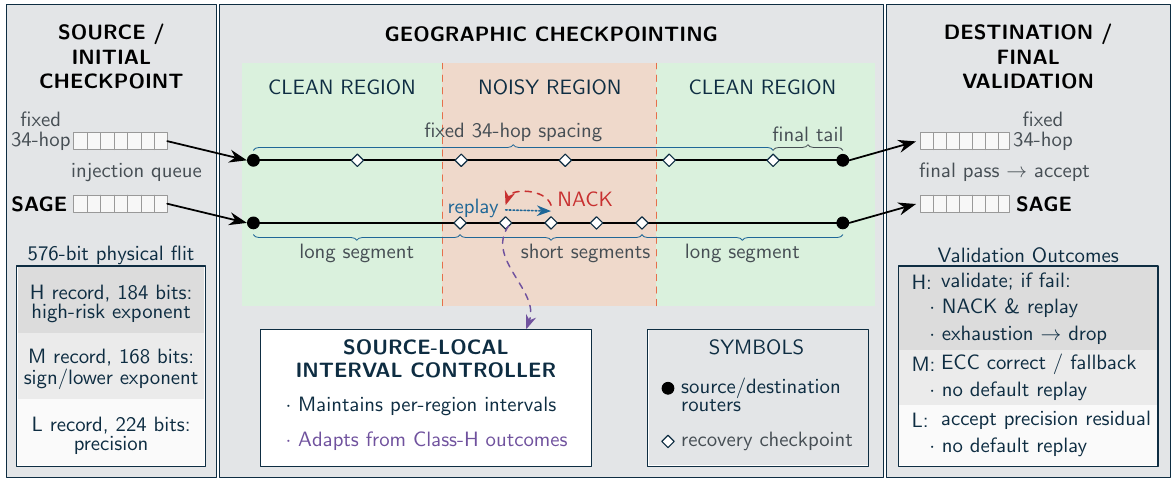}%
    \caption{
        \sage separates semantic replay eligibility from geographic replay distance. 
        Detected Class-\clH{} failures may trigger a protected NACK and full-flit replay. Class-\clM{} and Class-\clL{} outcomes do not trigger default network replay. 
        A source-local region table shortens future intervals after Class-\clH{} failures and gradually lengthens them after sustained clean windows.        
    }
    \Description{
        Overview diagram showing Class-H, Class-M, and Class-L records from source to destination.
        The fixed-interval architecture spaces checkpoints uniformly, whereas SAGE adapts checkpoint spacing to error geography.
    }
    \label{fig:sage-overview}
\end{figure*}

We present \sage, a semantic-aware geographic error-recovery architecture that separates two questions: \emph{whether a detected fault merits replay, and where that replay restarts}.
Prior approximate NoCs such as DEC-NoC suppress retransmission for low-order errors~\cite{chen2018decnoc}, while fault-tolerant NoCs adapt routing and recovery to physical faults~\cite{park2006faulttolerant}.
\sage combines a workload-calibrated semantic contract with source-local geographic checkpointing: numerical risk governs replay eligibility, while fault geography governs recovery distance.
This separation lets noisy regions use shorter recovery segments without imposing the same checkpoint spacing on clean regions.

The contract first enforces safety.
For the evaluated BF16-like data, Class-\clH{} contains high-significance exponent bits whose escaped errors may exceed a calibrated numerical envelope; Class-\clM{} contains sign and lower-exponent bits with important but bounded effects; and Class-\clL{} contains precision bits.
Policies must satisfy the Class-\clH{} silent-delivery risk budget before performance is considered.
Detected Class-\clH{} failures may trigger replay, whereas Class-\clM{} and Class-\clL{} outcomes do not trigger default network replay.

Because latency metrics alone can artificially reward policies that abandon recoverable packets early, \sage ranks admissible policies by \emph{quality-normalized terminal latency}: mean latency divided by delivered semantic quality (Section~\ref{sec:safety-before-performance}).
We also report latency and quality separately in our evaluation.

Our main comparison is against fixed 34-hop recovery in ten-seed direct-Garnet experiments.
At the stable synthetic operating point, with base bit-error rate (BER) $3\times10^{-5}$, \sage lowers mean latency by 28.0\% and all-outcome p99.5 latency by 36.0\%, while improving delivered semantic quality.
Together, these gains reduce quality-normalized terminal latency by 30.1\% (Table~\ref{tab:sage-direct-synthetic-3e5}).
All ten traffic- and seed-paired p99.5 comparisons favor \sage.

Under higher-BER synthetic stress at the same offered load, \sage maintains bounded queues while the fixed baseline accumulates backlog.
Application-derived DeiT-S traces and Garnet-calibrated wrapper traces also show lower mean and p99.5 latency at the evaluated BERs.

\sage implements geographic recovery with a small table of checkpoint intervals at each source (Figure~\ref{fig:sage-overview}).
Class-\clH{} failures shorten intervals in noisy regions; sustained clean windows allow longer intervals.
The source inserts a checkpoint when a segment reaches the shortest interval encountered since the preceding checkpoint.
Validation occurs only at segment endpoints.
A detected Class-\clH{} failure returns a protected negative acknowledgement to the preceding checkpoint and replays the complete physical flit, refreshing its \clH{}, \clM{}, and \clL{} contents.

We implement the endpoint and replay path in gem5 Garnet~\cite{binkert2011gem5,agarwal2009garnet} and synthesize a fully pipelined Class-\clH{}/Class-\clM{} receiver checker with initiation interval one.
Ordered per-VC staging and packet-specific credit ownership prevent younger bypass packets from overtaking packets undergoing validation.
Garnet executes a causally generated checkpoint-plan trace; the checker RTL is synthesized separately.

A complementary Garnet-calibrated wrapper isolates semantic replay eligibility: \sage{}+M retains \sage{}'s geographic checkpoint plan but also allows Class-\clM{} failures to trigger replay.
At base BER $3\times10^{-5}$, ten-seed results under the same finite retry budget show that this change multiplies retry DATA flit-hops per original by $2.30$ and raises drop rate from 7.42\% to 21.64\%, reducing delivered semantic quality by 8.3\%.
Thus, replaying bounded faults can sacrifice useful deliveries rather than improve delivered quality.
Separate decoder and HBM studies qualify Class-\clH{} silent-delivery risk and test portability (Section~\ref{sec:evaluation}).

This paper makes three contributions:
\begin{itemize}
    \item \textbf{A semantic reliability contract.}
    We separate catastrophic Class-\clH{} faults from bounded Class-\clM{}/Class-\clL{} damage and rank admissible policies by quality-normalized terminal latency.

    \item \textbf{A geographic checkpoint and replay architecture.}
    A source-local region table localizes short recovery segments to noisy geography; endpoint validation and full-flit replay preserve the semantic contract.

    \item \textbf{Implementation and quality-aware evaluation.}
    A Garnet protocol implementation and fully pipelined checker support experiments that quantify latency--quality tradeoffs, evaluate queue stability under fault stresses, and test application-derived traffic.
\end{itemize}

To support reproducibility, the complete gem5 Garnet protocol implementation, synthesizable checker RTL, and application-derived traces will be open-sourced upon acceptance.

%% file: sections/contract.tex
\section{Semantic Reliability Contract}
\label{sec:semantic-contract}

\sage separates the numerical consequence of a fault from the physical mechanism used to recover it.
The contract has two parts: a hard safety rule for catastrophic residuals and a quality--latency objective for bounded residuals.
The learning model calibrates class boundaries and quality weights offline; it is not evaluated per packet in the NoC control loop.

\subsection{Semantic Protection Classes}
\label{sec:semantic-classes}

For the BF16-like packets used in our NoC experiments, each 16-bit value is divided as
\begin{equation}
    \begin{aligned}
    \mathcal{H}&=\{e_4, e_5, e_6, e_7\},    \\
    \mathcal{M}&=\{s, e_0, e_1, e_2, e_3\}, \\
    \mathcal{L}&=\{f_0,\ldots,f_6\}.
    \end{aligned}
    \label{eq:sage-hml-split}
\end{equation}
Rather than relying on a rigid, universal bit index, \sage calibrates exact class boundaries offline to adapt to specific workload, layer, and numerical-format requirements.
A practical deployment first freezes a tensor envelope from clean statistics---for example, a six-standard-deviation envelope when that summary is meaningful, or an equivalent empirical tail quantile for non-Gaussian tensors.
A position belongs to Class-\clH{} when a plausible escaped residual, especially an exponent $0\to1$ transition, can create a non-finite value, leave that envelope, dominate a downstream reduction, or corrupt persistent state.

For the evaluated AlexNet distributions~\cite{krizhevsky2012alexnet}, this criterion places $e_4$--$e_7$ in Class-\clH{} for the dangerous upward direction, while $e_0$--$e_3$ remain finite and range bounded and are treated as Class-\clM{}; other models and tensors must be recalibrated.
Class-\clM{} contains faults that can materially change sign or scale but remain finite and range bounded under calibration.
Class-\clL{} contains precision faults whose expected effect is closer to quantization noise.
Appendix~\ref{app:exponent-recentering} illustrates optional transport-only exponent recentering, but no benefit from exponent recentering is credited in our results.

The \clH{}/\clM{}/\clL{} classification determines the default terminal action:
\begin{itemize}
    \item \textbf{Class-\clH{}:} Receives strong correction and detection, range/non-finite validation, and a safe terminal action such as replay, replacement, or drop.
    \item \textbf{Class-\clM{}:} Receives correction or detection but skips default network replay; bounded ambiguity invokes clipping, erasure, or approximate acceptance.
    \item \textbf{Class-\clL{}:} Uses opportunistic high-rate protection, truncation, erasure, or approximate acceptance and normally never triggers replay.
\end{itemize}

\subsection{Safety Before Performance}
\label{sec:safety-before-performance}

Let \(\pi\) denote a protection, checkpoint, and terminal-action policy under component error condition \(P\).
The policy is admissible only if its per-original-item Class-\clH{} catastrophic-risk probability satisfies
\begin{equation}
    P_{\mathrm{cat},H}(\pi,P)\le \epsilon_{\mathrm{cat}}.
    \label{eq:sage-catastrophic-constraint}
\end{equation}
The probability in (\ref{eq:sage-catastrophic-constraint}) includes correction, detection, range checking, all permitted replays, and the final safe action.
A policy that violates this constraint is rejected even when its average latency or raw goodput is attractive.

For an original item \(i\), let \(q_i\in[0,1]\) denote the calibrated semantic quality of its final accepted value.
Dropped items contribute zero, giving delivered quality
\begin{equation}
    q_{\mathrm{del}}(\pi,P)
    =\mathbb{E}\!\left[q_i\,\mathbf{1}_{\{i\ \mathrm{delivered}\}}\right]
    =(1-P_{\mathrm{drop}})q_{\mathrm{acc}}.
    \label{eq:sage-qdel}
\end{equation}

Among policies satisfying Eq.~\eqref{eq:sage-catastrophic-constraint}, we rank them by quality-normalized terminal latency:
\begin{equation}
    \begin{aligned}
    \Psi_{\rm del}(\pi,P)
        &= \frac{L_{\rm mean}(\pi,P)}
        {q_{\rm del}(\pi,P)}, \\
    \pi^\star 
        &= \operatorname*{arg\,min}_{\pi:\,P_{\rm cat,H}(\pi,P)\le\epsilon_{\rm cat}}
        \Psi_{\rm del}(\pi,P).
\begin{comment}        
    \pi^\star
        &= \arg\min_{\pi:\,
        P_{\rm cat,H}(\pi,P)\le\epsilon_{\rm cat}}
        \Psi_{\rm del}(\pi,P).
\end{comment}
    \end{aligned}
    \label{eq:quality-service}
\end{equation}
Here, $\pi^\star$ denotes an admissible policy minimizing $\Psi_{\rm del}$ at error condition $P$, and $L_{\rm mean}$ is the mean latency over all original items, measured from generation to final acceptance or retry-exhaustion detection.
\begin{comment}
Here, $L_{\rm mean}$ is the mean latency over all original items, measured from generation to final acceptance or retry-exhaustion detection. 
\end{comment}
It therefore captures source queueing, network traversal, checkpoint validation, replay, and terminal failure resolution.
Equation~\eqref{eq:quality-service} weighs recovery delay against delivered quality, including losses from drops and residual damage; we also report latency, drop rate, and $q_{\rm del}$ separately.

\subsection{Learning-Aware Calibration}
\label{sec:learning-aware-calibration}

The paper's learning law provides a compact calibration map from delivered semantic quality to task-level utility,
\begin{equation}
    A(p_{\mathrm{eff}})
    =A_{\infty}\!\left[1-K_1(1+p_{\mathrm{eff}})^{-K_2}\right],
    \qquad
    p_{\mathrm{eff}}=p \cdot q_{\mathrm{del}},
    \label{eq:sage-learning-law}
\end{equation}
where \(p\) is a normalized clean-data coordinate.
We fit (\ref{eq:sage-learning-law}) offline using clean learning curves and use targeted fault injection to calibrate the bounded Class-\clM{}/Class-\clL{} quality weights and the Class-\clH{} envelope.
The NoC controller does not evaluate this learning curve per packet.
Its online state is limited to geographic checkpoint statistics; the semantic class map and safety threshold are deployment parameters.

\subsection{Architectural Invariants}
\label{sec:sage-invariants}

The contract imposes four invariants on the architecture:
\begin{enumerate}
    \item an accepted Class-\clH{} residual must satisfy the deployment safety budget;
    \item a detected Class-\clH{} failure may trigger segment replay or safe drop, whereas Class-\clM{} and Class-\clL{} outcomes do not trigger default network replay;
    \item a Class-\clH{} replay retransmits the complete physical flit, thereby refreshing all semantic classes; and
    \item checkpoint placement may change recovery distance, but it must not weaken endpoint validation or packet ordering.
\end{enumerate}
%Section~\ref{sec:sage-architecture} describes the source-local geographic controller, endpoint checker, replay protocol, and ordered flow-control implementation that realize these invariants.
%\input{figures/sage_microarchitecture}

%% file: sections/architecture.tex
\section{\sage Architecture}
\label{sec:sage-architecture}

\begin{comment}
Figure~\ref{fig:sage-overview} summarizes the architecture.
\sage combines two decisions that conventional retry systems usually conflate: numerical semantics determine \emph{whether} an error may consume network replay, while spatial error history determines \emph{how far} replay must travel.
The design contains four mechanisms: a semantic physical-flit contract, a source-local geographic checkpoint controller, endpoint validation and segment replay, and an ordered flow-control path for mixing checked and ordinary traffic.
\end{comment}

Figure~\ref{fig:sage-overview} summarizes \sage, combining two decisions that conventional systems conflate: numerical semantics govern replay eligibility, while spatial history governs recovery distance. 
The design comprises four mechanisms: a semantic flit contract, a geographic checkpoint controller, endpoint validation and replay, and an ordered flow-control path for mixed traffic.

\subsection{Semantic Physical-Flit Contract}
\label{sec:sage-flit-contract}

The evaluated NoC carries one widened physical flit per source item,
\begin{equation}
    184_{H}+168_{M}+224_{L}=576\ \text{bits}.
    \label{eq:sage-flit-layout}
\end{equation}
The Class-\clH{} record contains 128 semantic bits, CRC32, and 24 BCH parity bits~\cite{bose1960bch}, forming shortened BCH$(184,160,t=3)$.
Class-\clM{} contains 160 semantic bits and eight BCH parity bits, forming shortened BCH$(168,160,t=1)$.
Class-\clL{} contains 224 uncoded precision bits.
For the fixed BF16-like mapping in Eq.~\eqref{eq:sage-hml-split}, no per-packet class mask is required; optional descriptors are protected as Class-\clH{} metadata.

Ordinary routers forward the flit without semantic decoding.
A selected recovery endpoint runs the Class-\clH{} and Class-\clM{} checkers in parallel.
The evaluated Class-\clH{} path performs BCH correction, CRC32 verification, and a registered range/non-finite check.
The Class-\clM{} path performs lightweight single-error correction and bounded fallback.
The Class-\clL{} field is delivered with its residual precision damage.
Thus, protection strength and terminal action differ within one physical transfer, but replay always refreshes the complete flit.

\subsection{Source-Local Geographic Checkpointing}
\label{sec:sage-geographic-controller}

Each source $s$ stores an interval $n_s(r)$ per error region.
Along receiving-router regions $r_1,\ldots,r_h$, with preceding checkpoint $c$, the next checkpoint is the first $j>c$ satisfying $j-c \ge \min_{c<u\le j} n_s(r_u)$, or the destination.
Region boundaries are not forced checkpoints; after placement, $c$ advances to $j$, so the minimum is recomputed over the next segment.
We call this region-local policy simply \sage; the conservative path-wide variant \sagep instead uses the smallest regional interval on the route for every segment.

The controller learns from Class-\clH{} outcomes over fixed windows.
A segment's feedback is apportioned to its traversed regions in proportion to the segment hops spent in each region.
For source-region pair $(s,r)$, an exponentially weighted score combines Class-\clH{} retries and terminal drops,
\begin{equation}
    z_{s,r}\leftarrow(1-\alpha)z_{s,r}
    +\alpha\frac{w_R R_{s,r}+w_D D_{s,r}}{E_{s,r}},
    \label{eq:sage-region-score}
\end{equation}
where $E_{s,r}$ is the observed hop exposure and drops receive the larger weight.
Let $c_{s,r}$ count low-score windows since the last high-score response or interval relaxation.  
If $z_{s,r}\ge\theta_{\rm high}$, the interval is halved and $c_{s,r}$ is reset.  
If $z_{s,r}<\theta_{\rm low}$, $c_{s,r}$ is incremented; after four such windows, the interval is increased by five hops and the counter is reset.
Scores between the two thresholds leave both states unchanged.

The interval update is
\begin{equation}
    n_s(r)\leftarrow
    \begin{cases}
        \max\!\left(n_{\min},\left\lceil n_s(r)/2\right\rceil\right),
            &z_{s,r}\ge\theta_{\mathrm{high}},\\
        \min\!\left(n_{\max},n_s(r)+5\right),
            &c_{s,r}\ge4,\\
        n_s(r),&\text{otherwise}.
    \end{cases}
    \label{eq:sage-table-update}
\end{equation}
The evaluated $100\times100$ configuration uses $n_0=67$, $n_{\min}=10$, $n_{\max}=199$, a 40K-packet update window, $\theta_{\mathrm{high}}=10^{-3}$, and $\theta_{\mathrm{low}}=10^{-4}$.
Halving responds quickly to a noisy region, whereas additive relaxation avoids immediately lengthening a recently hot segment.
At packet injection, the source snapshots the resulting checkpoint plan; later table updates affect only subsequent packets.

\subsection{Endpoint Validation and Segment Replay}
\label{sec:sage-endpoint-replay}

A checkpoint segment $k$ accumulates link errors and decodes only once at its selected endpoint. 
If $\mathbf{e}_{\ell}$ is the error mask contributed by link $\ell$, the checker observes
\begin{equation}
  \mathbf{e}_k = \bigoplus_{\ell \in k} \mathbf{e}_{\ell}.
  \label{eq:segment-error}
\end{equation}
Repeated flips to the same bit may cancel, and the replay distance is exactly the segment length.

A successful Class-\clH{} check commits the packet after the parallel Class-\clM{} result is available.
A Class-\clM{} beyond-radius result invokes its calibrated bounded action but does not generate a network replay in \sage.
A detected Class-\clH{} failure sends a protected NACK to the preceding checkpoint.
The preceding checkpoint retains the flit and sequence metadata while the segment remains unresolved and replays the segment upon receipt of a NACK.
The evaluated policy permits one original attempt plus at most two replays; retry exhaustion invokes a safe terminal drop.
Because the full 576-bit flit is resent, a Class-\clH{}-triggered replay also refreshes Class-\clM{} and Class-\clL{} without allowing those classes to independently drive congestion.

The evaluated timing has a 3-cycle ordinary router-plus-link hop.
The parallel Class-\clH{} path sets a 12-cycle validation surcharge, so a reached checkpoint hop costs 15 cycles all-in; the 5-cycle Class-\clM{} path adds no second surcharge.
These parameters are grounded by the standalone checker RTL and are separated from network queueing in the evaluation.

\subsection{Ordered VC Ownership and Credit Accounting}
\label{sec:sage-vc-ownership}

Endpoint checking must coexist with ordinary bypass traffic on the same input VC.
Holding the upstream VC for the full checker latency would serialize traffic, so the implementation may return an early free credit when a checked flit moves into bounded checker staging.
Doing so safely requires packet-specific ownership rather than an anonymous per-VC credit count.

For each input VC, \sage maintains an ordered owner queue.
Every checked or bypass arrival receives a monotonically increasing sequence record.
A younger bypass flit is placed in deferred ingress whenever an older checked owner remains unresolved, and it may enter the ordinary Garnet VC only after all older owners have committed or failed and the VC is idle.
Checker completion is likewise committed only from the head of the owner queue.

Each checked owner also carries an explicit early-credit token.
On successful reinsertion and later departure, the ledger suppresses the corresponding second credit exactly once.
On checker failure, NACK generation, or terminal drop, the same token is retired with that packet and cannot be charged to a younger flit.
All checker and deferred-ingress structures are bounded and apply backpressure when full.
These rules preserve per-VC packet order, prevent bypass traffic from overtaking a checked packet, and maintain exact credit ownership without sacrificing checker pipelining.

%% file: sections/implementation.tex
\input{figures/sage_microarchitecture}

\section{Implementation}
\label{sec:implementation}

This section describes the reference checkpoint controller, the direct-Garnet endpoint and replay implementation, and the synthesizable receiver-checker RTL.
Direct-Garnet experiments evaluate \sage and fixed 34-hop recovery, with \sagep as a checkpoint-placement ablation.
These policies share the network and protocol configuration; fixed 34-hop and \sage also share the executable and traffic trace.
The separate Garnet-calibrated wrapper evaluates all seven policies summarized in Table~\ref{tab:noc-summary}, with headline results in Table~\ref{tab:sage-wrapper-headline} and the complete sweep in Appendix~\ref{app:supplementary}.

\subsection{Reference Controller and Trace Interface}
\label{sec:implementation-controller}

The source-local controller is implemented in Python using the update rule in Eq.~\eqref{eq:sage-table-update}.
For every original packet it computes the deterministic route, snapshots the current checkpoint endpoints, and emits a trace containing generation time, packet ID, source, destination, and policy-independent workload metadata.
A companion checkpoint file records the internal endpoint positions for \sage and \sagep; fixed-interval policies derive their endpoints directly.

The direct-Garnet geographic policies use a \emph{trace-locked}, causally generated checkpoint sequence.
Offline, the controller processes the complete trace in packet-ID order and applies every 40K-packet update from expected Class-\clH{} retry/drop feedback only to later packets.
Garnet executes the immutable sequence without running the controller or feeding back realized queue or fault outcomes.
For DeiT-S, generation uses matching traffic, rate, and BER, not a uniform-random or higher-BER surrogate.
Thus, trace locking preserves reproducibility while isolating network execution from online controller overhead.

\subsection{Direct Garnet Protocol}
\label{sec:implementation-garnet}

We extend a pinned gem5 Garnet tree with one recovery controller adjacent to every router.
The evaluated configuration is a $100\times100$ mesh with 576-bit links, 2-cycle routers, and 1-cycle links.
Original and replay flits share an ordered DATA virtual network; protected NACK and lifecycle messages use a separate ordered CTRL virtual network.
DATA and CTRL use deterministic XY and YX routing, respectively.
The main configuration uses four VCs per vnet, eight-entry DATA and CTRL buffers, and a configured checker capacity of 16 entries per physical DATA input, including four checked-output entries.

CTRL receives bounded priority at the network interface and switch allocator: at most four consecutive CTRL grants are allowed while DATA is waiting.
Replay DATA has the same priority as original DATA.
Both vnets share the physical routers, crossbars, and links, so NACK and replay traffic contribute real contention and backpressure.
All messages are one physical flit in the evaluated model.
Transaction identifiers, endpoint positions, attempt numbers, and recovery bookkeeping are modeled as metadata outside the 576-bit physical-flit budget; their packet-format overhead is not included.

At an ordinary router, a DATA flit bypasses semantic checking.
At an internal checkpoint or final destination, the input unit diverts the flit into the 12-cycle checker pipeline; each physical DATA input can accept one checked transaction per cycle when unstalled.
A successful internal check advances the transaction to the next segment; destination success accepts the original.
A detected Class-\clH{} failure deletes the failed speculative copy and injects one protected CTRL NACK only to the immediately preceding checkpoint. 
That checkpoint restores its complete 576-bit snapshot, increments the attempt number, and replays the failed segment along its deterministic XY route.
Each segment permits at most two replays; a third failed attempt triggers a safe drop.
Our primary direct profile uses ideal local checkpoint handoff while injecting NACK and replay traffic explicitly; the implementation also exposes explicit checkpoint-commit and final-ack messages as a separate sensitivity.

\subsection{Segment Fault and Checker Model}
\label{sec:implementation-fault-checker}

The simulator composes link BERs into segment-effective XOR BERs (Appendix~\ref{app:semantic-detail}) and samples endpoint error counts for the \clH{}/\clM{}/\clL{} fields.
The direct-Garnet checker preserves replay timing but classifies beyond-radius outcomes behaviorally rather than executing the algebraic BCH decoder.

We implement both blocking and fully pipelined Class-\clH{} and Class-\clM{} receiver checkers. 
Full pipelining reduces the initiation interval to 1 cycle for both classes, while preserving their respective baseline latencies. 
Compared to the blocking baseline, pipelining increases DC cell area by 40.3\% for Class-\clH{} and 11.6\% for Class-\clM{} (34.7\% combined), with post-route placed-cell areas reported in Appendix Table~\ref{tab:app-implementation-cost}. 
The modeled 12-cycle Class-\clH{} endpoint surcharge remains unchanged because the range and non-finite checks are separately registered. 
Rare bounded-distance miscorrections that pass the CRC are evaluated in our RTL-matching decoder campaign.

\subsection{Ordered Input-VC Integration}
\label{sec:implementation-ordered-vc}

Figure~2 shows the ordered per-VC integration from Section~\ref{sec:sage-vc-ownership}.
Packet-specific ownership permits early upstream credit release while preserving checked/bypass order and preventing duplicate credits.
The modified \texttt{InputUnit} assigns every DATA arrival a per-VC sequence record, provides one bounded no-credit ingress slot per VC when the checker is full, and admits a bypass flit to the ordinary VC only when it is the oldest unresolved owner.
The modified \texttt{SwitchAllocator} returns the identity of the departing flit, allowing the input unit to consume or cancel the exact packet's early-credit token.
Success, failure, NACK generation, checkpoint transfer, and token retirement all commit from the head of the ownership queue.
The ordered input path and protocol settings are shared by fixed 34-hop, \sagep, and \sage.

\subsection{Checker Realization and Recovery State}
\label{sec:implementation-state}

The receiver uses fully pipelined Class-\clH{} and Class-\clM{} checkers with initiation interval one; Section~\ref{sec:evaluation-safety-cost} reports their synthesized cost.
The interval table requires 100~B per source, and each active checkpoint stores one 72~B payload snapshot plus metadata.
Checker staging contains 16 entries per DATA input, including four checked outputs.
The interval-update state remains in software, and finite checkpoint-store area and timing are not synthesized.
RTL and Garnet self-checks cover checker timing and backpressure, NACK/replay/drop behavior, checked/bypass ordering, and exact early-credit cancellation (Appendix~\ref{app:repro:gates}).

%% file: figures/sage_microarchitecture.tex
% Figure 2 replacement, revision 2.
% Keep the existing \input{figures/sage_microarchitecture} in implementation.tex.
% Requires tikz with arrows.meta and calc (already loaded by main.tex).
% Text is set at its final size. Do NOT wrap this figure in resizebox/scalebox.
% Manual line breaks and natural-width nodes prevent unwanted hyphenation.
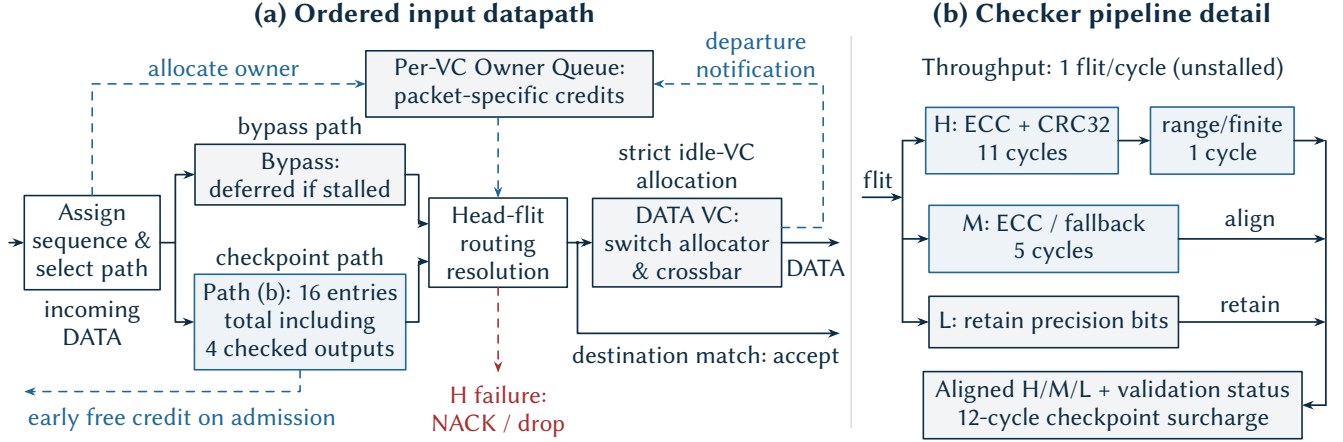
\begin{figure*}[t]
\centering
\begingroup
\definecolor{smarchInk}{RGB}{14,45,66}
\definecolor{smarchBlue}{RGB}{31,103,151}
\definecolor{smarchRed}{RGB}{165,48,48}
\definecolor{smarchPale}{RGB}{238,243,247}
\definecolor{smarchGray}{RGB}{243,244,245}
\begin{tikzpicture}[
    x=1cm,y=1cm,
    line cap=round,line join=round,
    every node/.style={
    font=\sffamily\fontsize{9.5}{10.5}\selectfont,
    text=smarchInk,align=center,inner sep=2pt,outer sep=0pt},
    smblock/.style={
    draw=smarchInk,fill=white,line width=.6pt,
    font=\sffamily\fontsize{9.5}{10.5}\selectfont,
    align=center,inner xsep=3pt,inner ysep=3pt,outer sep=0pt},
    smstate/.style={smblock,fill=smarchGray},
    smcheck/.style={smblock,fill=smarchPale,draw=smarchBlue},
    smdata/.style={-{Stealth[length=1.55mm,width=1.15mm]},
    draw=smarchInk,line width=.65pt},
    smctrl/.style={-{Stealth[length=1.5mm,width=1.1mm]},
    draw=smarchBlue,dashed,line width=.6pt},
    smfail/.style={smctrl,draw=smarchRed},
    smlabel/.style={
    font=\sffamily\fontsize{9.5}{10.5}\selectfont,
    text=smarchInk,align=center,inner sep=1pt,outer sep=0pt},
    smtitle/.style={
    font=\sffamily\bfseries\fontsize{10.5}{11.5}\selectfont,
    text=smarchInk,align=center,inner sep=0pt,outer sep=0pt}
]

% 17.66 cm canvas fits the 17.78 cm text width of acmart sigplan.
% There is no scaling of coordinates, nodes, or text.
\path[use as bounding box] (-.04,-.15) rectangle (17.62,5.66);
\draw[black!20,line width=.4pt] (11.12,.00) -- (11.12,5.18);
\node[smtitle] at (5.48,5.46) {(a) Ordered input datapath};
\node[smtitle] at (14.44,5.46) {(b) Checker pipeline detail};

% (a) One physical DATA input; ownership/deferred ingress are per VC.
\node[smblock,minimum width=1.78cm,minimum height=1.10cm]
    (in) at (1.08,2.45) {Assign\\sequence \&\\select path};
\node[smstate,minimum width=2.78cm,minimum height=.78cm]
    (bypass) at (3.83,3.35) {Bypass:\\deferred if stalled};
\node[smcheck,minimum width=2.78cm,minimum height=1.16cm]
    (check) at (3.83,1.40)
    {Path (b): 16 entries\\total including\\4 checked outputs};
\node[smblock,minimum width=1.83cm,minimum height=1.10cm]
  (head) at (6.45,2.45) {Head-flit\\routing\\resolution};
\node[smstate,minimum width=2.50cm,minimum height=1.10cm]
  (vc) at (8.95,2.45) {DATA VC:\\switch allocator\\\& crossbar};
\node[smstate,minimum width=3.80cm,minimum height=.82cm]
  (owners) at (6.60,4.55)
  {Per-VC Owner Queue:\\packet-specific credits};

% Payload fork. Each arrow terminates at the actual node boundary.
\draw[smdata] (-.02,2.45) -- (in.west);
\node[smlabel,anchor=north] at ([yshift=-3pt]in.south)
  {incoming\\DATA};
\coordinate (fork) at (2.18,2.45);
\draw[draw=smarchInk,line width=.65pt] (in.east) -- (fork);
\draw[smdata] (fork) |- (bypass.west);
\draw[smdata] (fork) |- (check.west);
\node[smlabel] at (3.83,3.96) {bypass path};
\node[smlabel] at (3.83,2.23) {checkpoint path};
\draw[smdata] (bypass.east) -- (5.40,3.35)
  |- ($(head.west)+(0,.25)$);
\draw[smdata] (check.east) -- (5.40,1.40)
  |- ($(head.west)+(0,-.25)$);
\draw[smdata] (head.east) -- (vc.west);
\node[smlabel,anchor=south] at ([yshift=3pt]vc.north)
  {strict idle-VC\\allocation};
\draw[smdata] (vc.east) -- (10.97,2.45);
\node[smlabel] at (10.65,2.12) {DATA};

% Destination success is a branch BEFORE VC forwarding.
%\draw[smdata] (7.50,2.45) -- (7.50,3.51) -- (10.05,3.51);
\draw[smdata] (7.50,2.45) -- (7.50,1.25) -- (10.97,1.25);
\fill[smarchInk] (7.50,2.45) circle[radius=.65pt];
\node[smlabel,anchor=south] at (9.20,0.75)
  {destination match: accept};

% Owner allocation, in-order resolution, and identity-aware credit feedback.
\draw[smctrl] (in.north) |- (owners.west);
\node[smlabel,text=smarchBlue,anchor=south] at (2.82,4.62)
  {allocate owner};
\draw[smctrl] (owners.south -| head.north) -- (head.north);
\draw[smctrl] ($(vc.east)+(0,.20)$) -- (10.75,2.65)
  -- (10.75,4.55) -- (owners.east);
\node[smlabel,text=smarchBlue,anchor=south] at (9.86,4.61)
  {departure\\notification};

% Return an upstream free credit only after admission to bounded staging.
\draw[smctrl] (check.south) -- (3.83,.48) -- (.18,.48);
\node[smlabel,text=smarchBlue,anchor=north] at (2.27,.30)
  {early free credit on admission};

% Failed checked owners resolve in order; only H invokes network recovery.
\draw[smfail] (head.south) -- (6.45,.73);
\node[smlabel,text=smarchRed,anchor=north] at (6.45,.59)
  {H failure:\\NACK / drop};

% (b) Parallel H/M processing and retained Class-L bits.
\node[smlabel] at (14.45,4.75)
  {Throughput: 1 flit/cycle (unstalled)};
\node[smcheck,minimum width=2.45cm,minimum height=.83cm]
  (h) at (13.38,3.80) {H: ECC + CRC32\\11 cycles};
\node[smcheck,minimum width=1.90cm,minimum height=.83cm]
  (range) at (16.03,3.80) {range/finite\\1 cycle};
\node[smcheck,minimum width=3.30cm,minimum height=.80cm]
  (m) at (13.80,2.50) {M: ECC / fallback\\5 cycles};
\node[smstate,minimum width=3.30cm,minimum height=.68cm]
  (l) at (13.80,1.40) {L: retain precision bits};
\node[smstate,minimum width=5.01cm,minimum height=.75cm]
  (out) at (14.58,.30)
  {Aligned H/M/L + validation status\\12-cycle checkpoint surcharge};

% Logical field fan-out and gather. Vertical buses stay OUTSIDE all boxes.
\coordinate (split) at (11.80,3.05);
\draw[draw=smarchInk,line width=.65pt]
  (split |- l.west) -- (split |- h.west);
\draw[smdata] (11.27,3.05) -- (split);
\node[smlabel] at (11.48,3.28) {flit};
\draw[smdata] (split |- h.west) -- (h.west);
\draw[smdata] (h.east) -- (range.west);
\draw[smdata] (split |- m.west) -- (m.west);
\draw[smdata] (split |- l.west) -- (l.west);
\coordinate (gather) at (17.40,.30);
\draw[draw=smarchInk,line width=.65pt]
  (gather) -- (gather |- range.east);
\draw[smdata] (range.east) -- (gather |- range.east);
\draw[smdata] (m.east) -- (gather |- m.east);
\draw[smdata] (l.east) -- (gather |- l.east);
\draw[smdata] (gather) -- (out.east);
\node[smlabel] at (16.40,2.74) {align};
\node[smlabel] at (16.40,1.64) {retain};
\end{tikzpicture}
\endgroup
  \caption{Ordered input integration and checker pipelines in SAGE.
  (a) One physical DATA input is shown; ownership and deferred ingress are per VC.
  Checked and bypass transactions resolve in owner order.
  An early-credit token prevents a second credit on departure and retires on failure.
  H failure invokes a NACK to the preceding checkpoint or safe drop at exhaustion.
  (b) H and M run in parallel; H sets the 12-cycle surcharge.
  Solid/dashed arrows denote data/control.
  Checker RTL is synthesized separately from the modeled ownership/credit integration.}
  \ifdefined\Description
    \Description{Two-panel logical microarchitecture diagram. An incoming DATA
    flit receives a per-VC sequence record, then follows either the bypass path
    or the bounded checker path. Both paths are resolved in per-VC owner order.
    Success forwards through an idle DATA VC or accepts at the destination;
    Class-H failure generates a NACK or safe drop. A packet-specific credit
    ledger records early credits and departing identities. The checker inset
    shows the 11-cycle Class-H ECC and CRC32 pipeline followed by a one-cycle
    range and finite-value check, in parallel with the five-cycle Class-M
    ECC/fallback path and the retained Class-L bits.}
  \fi
  \label{fig:sage-microarchitecture}
\end{figure*}

%% file: sections/methodology.tex
\section{Experimental Methodology}
\label{sec:methodology}

We deploy a multi-tiered evaluation framework, leveraging complementary simulators to isolate different architectural mechanisms.
The Garnet-calibrated wrapper explores policies and stress points efficiently; direct Garnet, built on gem5 Garnet~\cite{agarwal2009garnet}, exposes persistent queues and explicit recovery traffic under both uniform-random and application-derived traffic; the endpoint decoder campaign qualifies rare Class-\clH{} escapes; and Ramulator~2.0~\cite{luo2024ramulator2} tests whether the semantic recovery rule transfers to HBM.
Appendix~\ref{app:repro} consolidates the campaign contracts, common configuration, stationarity tests, and acceptance gates.

\subsection{Common NoC Configuration and Baselines}
\label{sec:methodology-common-noc}

Table~\ref{tab:noc-summary} summarizes the NoC setup and wrapper policies~\cite{benini2002noc,agarwal2009garnet}.
All policies share the BF16-like representation and codeword budget: semantic protection~\cite{masnick1967linear,albanese1996priority,borade2009uep,huang2020functional} is evaluated at fixed precision~\cite{gupta2015limitedprecision,reagen2016minerva,micikevicius2022fp8,rouhani2023microscaling}.
Faults are conditionally independent given the fixed BER map; temporal and cross-link burst correlations are outside scope.
Appendix~\ref{app:repro:configuration} gives the full configuration and hot-region coordinates.

\begin{table}[t]
    \caption{
        NoC setup, traffic, and wrapper policies. Rates are in packets/router/cycle; direct-only settings are marked.
    }
    \label{tab:noc-summary}
    \centering
    \fontsize{9.5}{11}\selectfont
    \setlength{\tabcolsep}{3pt}
    \renewcommand{\arraystretch}{1.03}
    \begin{tabularx}{\columnwidth}{@{}>{\raggedright\arraybackslash}p{0.29\columnwidth}>{\raggedright\arraybackslash}X@{}}
        \toprule
        \textbf{Item} & \textbf{Evaluated setting} \\
        \midrule
        Network & $100\times100$ mesh; 576-bit links; XY DATA; 2-cycle routers, 1-cycle links \\
        Direct Garnet & YX CTRL; 4 VCs/vnet; 8 entries/VC for DATA and CTRL \\
        Recovery & 12-cycle checkpoint surcharge; one original plus at most two replays/segment \\
        Fault map & $10\times10$ regions of $10\times10$ routers; receiving-region BER; four hot regions at $100\times$ base BER \\
        Uniform-random & Wrapper/direct Garnet: offered 0.0108; clean saturation 0.027 \\
        DeiT-S (primary) & Immutable, rate-shaped communication trace; direct Garnet at rate 0.009 \\
        \midrule
        Fixed baselines & End-to-end, mean-path, fixed 34-hop, and fixed 10-hop intervals \\
        \sage & Region-local geographic checkpointing \\
        \sagep & Route-wide minimum regional interval \\
        \textsc{SAGE}+M & \sage's plan, with Class-\clM{} beyond-radius replay (negative control) \\
        \midrule
        Wrapper sampling & 10 common seeds; 6M warm-up, then 250K measured originals/policy/BER/seed \\
        \bottomrule
    \end{tabularx}
\end{table}

DEC-NoC~\cite{chen2018decnoc} is not a matched numerical baseline: it assumes per-value error thresholds, variable-length MSB protection, and optional integer-to-floating-point conversion.
The fixed intervals isolate recovery distance; \textsc{SAGE}+M isolates Class-\clM{} replay eligibility.

\subsection{Garnet-Calibrated Wrapper Sweep}
\label{sec:methodology-wrapper}

The wrapper XOR-accumulates BER along each planned segment, samples endpoint \clH{}/\clM{}/\clL{} outcomes, and maps replay flit-hops to effective load on the measured clean Garnet latency--throughput curve~\cite{agarwal2009garnet}.
It models neither persistent queues nor explicit NACK/replay packets, so it estimates load margin rather than final contention and needs no post-guard.
After warm-up, measurement counters reset; learned tables and 40K-packet updates continue.
The seven-policy sweep uses $P_{\rm base}\in\{2,3,4,5\}\times10^{-5}$; fixed 34-hop and \sage are paired at $\{2,3,5,10,20,50\}\times10^{-5}$ (Appendix~\ref{app:wrapper:matrix}--\ref{app:wrapper:stress}).
Points above $5\times10^{-5}$ are detected-failure stress tests outside the CRC-qualified envelope.

\subsection{Synthetic Traffic, Sampling, and Pairing}
\label{sec:methodology-direct}

\input{figures/fig_methodology_timeline}

Direct Garnet~\cite{agarwal2009garnet} consumes a trace-locked checkpoint sequence generated causally over the full traffic trace. It neither runs the controller nor updates $n_s(r)$ from realized outcomes.
For a given seed, fixed 34-hop and \sage share the uniform-random traffic trace, four-hot-region BER map, executable, and network configuration.

Figure~\ref{fig:methodology-timeline} shows the shared 17.25M-original timeline. The synthetic headline, paired $3\times10^{-4}$, and appendix-only $3\times10^{-3}$ campaigns report all 15.25M post-warm-up originals; $10^{-3}$ reports the 250K cohort, while DeiT-S reports the complete trace.
The 2M--6M prefix contributes controller and queue history to the synthetic headline; causal updates continue afterward.

At $P_{\rm base}=3\times10^{-5}$, we evaluate ten traffic- and seed-paired fixed 34-hop/\sage runs and ten independently validated \sagep ablation runs under the same topology, protocol settings, offered load, and measurement contract.
\sagep uses the route's smallest regional interval for every segment; paired confidence intervals remain limited to fixed 34-hop versus \sage.
Two main-text five-seed, traffic- and seed-paired stress campaigns compare fixed 34-hop and \sage at $P_{\rm base}=3\times10^{-4}$ and $10^{-3}$ under the same offered load.

An appendix-only $3\times10^{-3}$ campaign follows the same run contract and a final-five 5K-window diagnosis; both policies drop more than 93\% of originals.
All stress points lie outside the CRC-qualified operating envelope and test queue behavior at one load, not exact saturation thresholds.
For stationarity, $B(t)=N_{\rm gen}(\le t)-N_{\rm term}(\le t)$; we diagnose $3\times10^{-4}$ with the final five full-rate 10K-cycle windows plus a 5K-bin check and $10^{-3}$ over cycles 40K--150K plus the final 50K cycles.
Persistent completion below arrival with growing backlog denotes saturation; matched rates and bounded backlog denote queue stability, while drop rate separately determines useful service.

The endpoint oracle computes the segment-effective XOR BER and
deterministically keys each reached-endpoint sample by seed, packet ID, policy-dependent segment, attempt, checkpoint endpoint/incoming-link identity, and semantic class.
Because segmentation is policy-dependent, exact physical fault masks differ.
We therefore call fixed 34-hop versus \sage traffic- and seed-paired, not exact common-random-number paired.
Statistics use policy-dependent endpoint outcomes within each shared traffic seed.

For operating-margin diagnostics, we additionally report source-queue time, the final drain after traffic generation ends, and maximum directed-link demand.
Directed-link demand is the total post-warm-up flit traversals on a link divided by the number of post-warm-up cycles during which original traffic is still being generated.
Values below one are feasible on a one-flit-per-cycle link; values above one imply that the generated workload requires backlog accumulation on that link.

\subsection{Application-Derived DeiT-S Traffic}
\label{sec:methodology-deit}

To test whether the policy ordering persists beyond uniform-random traffic, we evaluate an immutable, rate-shaped communication trace derived from non-distilled DeiT-S/16 inference on selected ImageNet-1K validation images~\cite{touvron2021deit,russakovsky2015imagenet}.
The campaign uses 12 physical replicas with six in-flight contexts per replica and caps aggregate packet eligibility at 90 originals/cycle, equivalent to $r_{\rm DeiT}=0.009$ over the 10,000-router mesh.

A seed-1 rate scout swept the 0.0085–-0.010 range to locate the structured-traffic queueing knee, selecting 0.009 as the highest tested common rate before either conservative cross-cell showed abrupt injection overhang.
Appendix~\ref{app:deit-analysis} gives the diagnostic values.
The earlier 108-original/cycle setting ($r=0.0108$) is retained only at $P_{\rm base}=0$ as a high-load clean-channel diagnostic, not as a normal clean baseline or a point directly comparable with the lower-rate faulted runs.

At $r_{\rm DeiT}=0.009$, $P_{\rm base}=3\times10^{-6}$ is the application-derived operating point; $10^{-5}$ and $2\times10^{-5}$ test increasing fault stress.
These BERs are lower than in the uniform-random study because the stage-dependent DeiT-S bursts leave less recovery-traffic headroom; the unchanged $100\times$ multiplier gives hot-region BERs of $3\times10^{-4}$, $10^{-3}$, and $2\times10^{-3}$.
The application-derived and synthetic campaigns therefore test the same mechanism in different regimes, and their absolute latencies are not compared across workloads.

For each nonzero point, the controller causally generates the complete per-packet sequence from the matching DeiT-S trace, rate, and BER.
It emits packet $i$'s plan before accumulating its model-derived Class-\clH{} feedback, so each 40K-packet update affects only later IDs.
Direct Garnet executes the immutable sequence; the first 6M IDs (2M warm-up plus 4M pre-guard) provide controller and queue history, and adaptation continues through the cohort and following 11M originals.

The completed grid contains ten traffic- and seed-paired runs of fixed 34-hop and \sage at each of the three BERs.
Within each pair, packet IDs, routes, generation cycles, executable, BER map, and network configuration match; checkpoint plans and policy-dependent endpoint outcomes differ.
We report per-seed statistics over the complete 17.25M-original finite trace; Appendix~\ref{app:deit-trace} specifies the coverage of the 5K-cycle window and final-quiescence diagnostics.

Because the trace has no verified repetition period, they are paired finite-workload outcomes rather than steady-state or saturation estimates.
The experiment evaluates network-level contention, queueing, and recovery under application-derived communication; it does not execute end-to-end DeiT-S inference with corrupted tensor payloads.

\subsection{Metrics and Statistical Treatment}
\label{sec:methodology-statistics}

Primary latency spans generation to either final acceptance or retry-exhaustion detection over the campaign populations defined in Sections~\ref{sec:methodology-wrapper}--\ref{sec:methodology-deit}.
Terminal quantiles are computed separately for each seed with one fixed estimator applied identically to every policy.
Accepted-only latency is a sensitivity rather than the headline metric.
Latency quantiles from a nonstationary stress run are descriptive consequences of the growing queue rather than steady-state estimates.
We additionally report mean latency, source-queue and network components, retry DATA and CTRL flit-hops, planned and reached checkpoints, drop rate, $q_{\rm del}$, and quality-normalized terminal latency $\Psi_{\rm del}=L_{\rm mean}/q_{\rm del}$.

Statistics are computed per seed before aggregation; packets are never pooled across seeds to form a paper quantile.
For paired comparisons, we report the mean paired difference, a two-sided 95\% Student-$t$ confidence interval, and the number of pairs favoring \sage.
All traffic, checkpoint-plan, executable, BER-map, and configuration hashes are checked before a run enters analysis.
For the application-derived campaign, the rate-cap audit and input-provenance hashes are checked as well.

\subsection{Calibration, Decoder Qualification, and HBM}
\label{sec:methodology-supporting}

We calibrate the semantic contract offline by fitting Eq.~\eqref{eq:sage-learning-law} to clean learning curves and comparing fit and held-out error against exponential and power-law baselines~\cite{hestness2017predictable,kaplan2020scaling,viering2022learningcurves}.
Targeted injections into exponent MSBs, lower-exponent bits in both directions, the sign bit, and magnitude-conditioned mantissa positions determine class boundaries and bounded-quality weights rather than NoC timing~\cite{li2017errorpropagation,mahmoud2020pytorchfi,reagen2018ares,rakin2019bitflip}.

The Class-\clH{} decoder qualification uses the ten-seed, 6M-warm-up \sage wrapper exposure at the four principal BERs.
For every reached checkpoint attempt, including retries, the exporter records the segment-effective BER.
Monte Carlo samples this exposure, injects IID flips into the shortened BCH/CRC record, and runs the RTL-matching decoder with locator roots outside transmitted positions rejected.
Five predeclared 10M-trial seeds per CRC32/CRC16 group give simultaneous one-sided 95\% Clopper--Pearson bounds over eight groups.
Per-original risk is conservatively bounded by $U_{\mathrm{pkt}}^{95}\leq \bar{A}U_{\mathrm{att}}^{95}$, where $\bar{A}$ is the mean number of reached Class-\clH{} attempts per original and $U_{\mathrm{att}}^{95}$ is the per-attempt upper bound. This composition assumes no independence between attempts and credits no range checks.
CRC16 reuses the same exposure and is not separately relearned.

A five-seed Ramulator~2.0 HBM3 pilot~\cite{luo2024ramulator2} compares Class-\clH{}-only and all-class recovery at 40\% of clean saturation. 
Appendix~\ref{app:hbm-portability} gives its configuration, accounting, and results.

%% file: figures/fig_methodology_timeline.tex
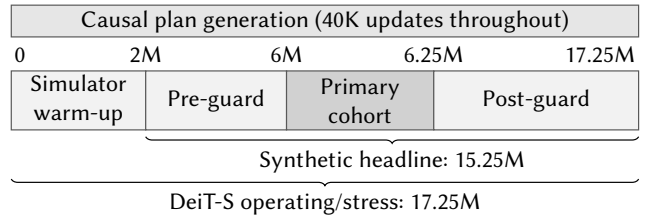
\begin{figure}[b]
    \centering
    \begin{tikzpicture}[
        x=\dimexpr\columnwidth/100\relax, y=1cm,
        font=\small\sffamily,
        every node/.style={inner sep=1pt, align=center},
        phase/.style={draw=black!65, line width=0.45pt},
        span/.style={decorate,
        decoration={brace, mirror, amplitude=2.5pt},
        line width=0.45pt}
    ]
    % All horizontal positions are fractions of one column.
    % Segment widths are schematic, not proportional to packet counts.
    \path[phase, fill=black!10] (1,1.23) rectangle (99,1.65);
%    \node[font=\scriptsize\sffamily] at (50,1.44)
%        {Causal plan generation (40K updates throughout)};
    \node at (50,1.44)
        {Causal plan generation (40K updates throughout)};

    % Packet-ID boundaries shared by plan generation and Garnet execution.
    \node[anchor=west] at (1,1.00) {0};
    \node at (22,1.00) {2M};
    \node at (44,1.00) {6M};
    \node at (67,1.00) {6.25M};
    \node[anchor=east] at (99,1.00) {17.25M};

    % Direct-Garnet injection sequence; each original is simulated once.
    \path[phase, fill=black!5]  (1,0)   rectangle (22,0.78);
    \path[phase, fill=black!5]  (22,0)  rectangle (44,0.78);
    \path[phase, fill=black!18] (44,0)  rectangle (67,0.78);
    \path[phase, fill=black!5]  (67,0)  rectangle (99,0.78);
    \node at (11.5,0.39)    {Simulator\\warm-up};
    \node at (33,0.39)      {Pre-guard};
    \node at (55.5,0.39)    {Primary\\cohort};
    \node at (83,0.39)      {Post-guard};

    % Braces identify reporting populations, not additional phases.
    \draw[span] (22,-0.09) -- (99,-0.09);
    \node[anchor=north] at (60.5,-0.23)
        {Synthetic headline: 15.25M};
    \draw[span] (1,-0.65) -- (99,-0.65);
    \node[anchor=north] at (50,-0.79)
        {DeiT-S operating/stress: 17.25M};
    \end{tikzpicture}
    \caption{
        Direct-Garnet timeline (packet IDs; not to scale).
        The controller precomputes a causal per-packet plan sequence; the $10^{-3}$ synthetic stress reports only the primary cohort.
    }
    \label{fig:methodology-timeline}
    % ACM accessibility description; harmless with other document classes.
    \Description{
        A shared packet-ID timeline from zero to 17.25 million.
        Across the complete trace, the offline controller generates checkpoint plans causally, updating every 40 thousand originals so each packet uses only feedback from earlier packets. Direct Garnet executes the resulting immutable sequence and partitions it into 2 million simulator warm-up originals, a 4 million pre-guard, a 250 thousand primary cohort, and an 11 million post-guard. Synthetic headline statistics use packet IDs from 2 million to 17.25 million; DeiT-S operating and fault-stress statistics use all originals. 
        The synthetic stress at base BER ten to the minus three reports only the primary cohort.
    }
\end{figure}

%% file: sections/evaluation.tex
\section{Evaluation}
\label{sec:evaluation}

We evaluate SAGE's semantic contract, latency--quality tradeoffs, and queue stability, then assess safety, implementation cost, and portability.

\subsection{Semantic Calibration Supports the Contract}
\label{sec:evaluation-semantic}

The calibration experiments support the two distinctions that \sage relies on: Class-\clH{} residuals require a hard safety rule, whereas Class-\clM{} and Class-\clL{} residuals can be ranked by bounded quality loss.
Across the five evaluated learning curves, Eq.~\eqref{eq:sage-learning-law} has lower MSE than the exponential-residual and power-law alternatives.
For example, on the DenseNet-121/ESC-50 mean curve its MSE is 0.144 squared percentage points, versus 1.468 and 0.602, and its held-out RMSE is 0.388 versus 2.633 and 1.541.
The learning law is therefore sufficiently accurate for the policy-ranking role in Section~\ref{sec:semantic-contract}; it is not used to predict online network state.

\paragraph{Does $q_{\rm del}$ track task accuracy?}
A curve fit alone does not show that a communication-derived quality factor tracks downstream task behavior.
We therefore test the mapping independently with the AlexNet/CIFAR-100 sign-bit experiment at normalized data ratio $p=2$.
Let $A_0(\cdot)$ denote the clean fitted learning curve in Eq.~\eqref{eq:sage-learning-law}.
The representation-level sign-flip energy anchor fixes $\lambda_s=4$, without refitting each noisy point, so that
\begin{equation}
    q_{\rm del}=q_{\rm acc}=\frac{1}{1+4\rho_s},
    \qquad
    A_{\rm pred}=A_0(2q_{\rm del}).
    \label{eq:sage-qdel-task-check}
\end{equation}
No item is dropped in this controlled experiment, hence $q_{\rm del}=q_{\rm acc}$.
Across five sign-bit BERs, the maximum absolute prediction error is 0.11 percentage points and the RMSE is 0.058 percentage points.
This validates the bounded-residual component of $q_{\rm del}$ in one measured task setting; the factor $(1-P_{\rm drop})$ in Eq.~\eqref{eq:sage-qdel} then assigns exactly zero quality to terminally dropped originals.

The five measured and predicted points are listed in Appendix Table~\ref{tab:app-qdel-task-check}.
Representative fault-injection evidence is summarized below and tabulated in Appendix Table~\ref{tab:app-semantic-evidence}.

Activating exponent-MSB faults in DenseNet-121 training drives all later checkpoints to the 2.0\% chance level; safe restoration recovers 83.0\% accuracy, close to the 83.3\% clean control.
By contrast, an upward lower-exponent fault is harmful but bounded: at BER $10^{-4}$, exponent bit $e_3$ reduces AlexNet top-1 accuracy to 63.38\%, whereas the matched downward direction retains 69.89\% and produces no NaN or infinity.
The evaluated $e_4$--$e_7$ boundary is a calibrated AlexNet case, not a universal partition: Class-\clH{} is assigned when a plausible upward residual crosses the tensor envelope described in Section~\ref{sec:semantic-classes}.
Mantissa-MSB damage remains smaller and magnitude dependent.
A fixed-budget sensitivity in Appendix~\ref{app:semantic-detail} promotes $e_3$, the sign bit, or both into Class-\clH{}; all tested variants remain safety-qualified, but none improves the optimistic fixed-latency estimate of $\Psi_{\rm del}$.

A matched 48-bit BCH-only allocation gives the same conclusion at the coding layer.
At a 198-hop envelope, H5--M1--L0 achieves a 95\% Class-\clH{} silent-delivery upper bound of $4.30\times10^{-8}$.  
By contrast, the equal H2--M2--L2 allocation yields an upper bound of $4.56\times10^{-3}$ and does not certify the $10^{-6}$ safety budget.
Neither result relies on CRC or range-check credit.
Thus, semantic allocation changes the feasible policy set before latency is considered.

\subsection{Ten-Seed Wrapper Design Space}
\label{sec:evaluation-wrapper}

The completed 6M-warm-up wrapper sweep isolates the architectural mechanism over more policies and BER points than are practical in packet-level simulation.
It distinguishes three possible sources of benefit: replaying only catastrophic failures, localizing short segments to noisy geography, and globally imposing a short interval.

The critical semantic negative control is \textsc{SAGE}+M: it preserves \sage's geographic checkpoint plan and changes only whether bounded Class-\clM{} failures may invoke replay.
The results use ten common seeds, 6M warm-up originals, and 250K measured originals per seed, matching Section~\ref{sec:methodology-wrapper}.

\begin{table}[t]
    \centering
    \caption{
        Garnet-calibrated wrapper results at $P_{\rm base}=3\times10^{-5}$ (ten common seeds; 6M warm-up and 250K measured originals per seed).
        $R_h$ is retry DATA flit-hops/original; P/C is planned segments/reached checks.  
        The complete sweep appears in Appendix~\ref{app:wrapper}.
    }
    \label{tab:sage-wrapper-headline}
    \small
    \setlength{\tabcolsep}{3pt}
    \begin{tabular}{@{}lrrrrrr@{}}
        \toprule
        Policy          & $R_h$             & P/C                   & $L_{\rm mean}$    
                        & $L_{99.5}$        & $q_{\rm del}$         & Drop              \\
        \midrule
        End-to-end      & 29.780            & 1.00/1.38             & 479.9             
                        & 2587.8            & 0.8018                & 0.1508            \\
        Mean-path       & 20.708            & 1.49/1.79             & 399.9             
                        & 1507.6            & 0.8106                & 0.1393            \\
        Fixed 34-hop    & 11.210            & 2.45/2.66             & 328.1             
                        & 1137.3            & 0.8263                & 0.1162            \\
        Fixed 10-hop    & 2.786             & 7.12/7.14             & 318.6             
                        & 785.8             & 0.8626                & 0.0610            \\
        \sagep          & 2.799             & 4.77/4.79             & 290.5             
                        & 773.8             & 0.8593                & 0.0609            \\
        \textbf{\sage}  & \textbf{2.980}    & \textbf{3.06/3.13}    & \textbf{270.9}    
                        & \textbf{726.6}    & \textbf{0.8504}       & \textbf{0.0742}   \\
        \textsc{SAGE}+M & 6.867             & 3.06/2.86             & 280.2             
                        & 944.2             & 0.7798                & 0.2164            \\
        \bottomrule
    \end{tabular}
\end{table}

\begin{figure}[t]
    \centering
    \input{figures/fig_tradeoff_q_latency}
    \caption{
        Quality--latency trade-off in the Garnet-calibrated wrapper at
        $P_{\rm base}=3\times10^{-5}$ (ten-seed means; Table~\ref{tab:sage-wrapper-headline}).
        Lower-right is better. Dashed lines show constant
        $L_{\rm mean}/q_{\rm del}$; filled markers identify non-dominated
        policies at the plotted means.
    }
    \Description{
        Scatter plot of delivered quality against mean terminal latency for
        seven wrapper policies at base BER 3e-5. SAGE has lower latency and
        higher quality than end-to-end, mean-path, fixed 34-hop, and SAGE+M.
        SAGE-P and fixed 10-hop deliver slightly higher quality at higher
        latency. Filled markers identify SAGE, SAGE-P, and fixed 10-hop as
        non-dominated at the plotted means. Dashed reference lines have
        latency-to-quality ratios of 320, 400, and 500 cycles.
    }
    \label{fig:quality_latency_tradeoff}
\end{figure}
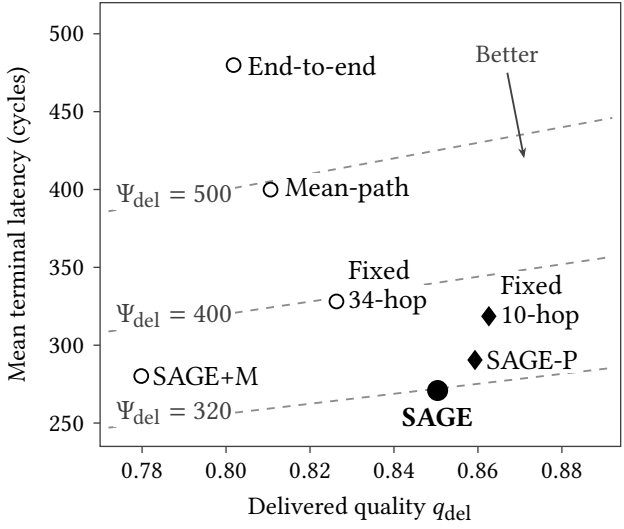

Table~\ref{tab:sage-wrapper-headline} exposes the mechanism at the headline point.  
Relative to the fixed 34-hop interval, \sage reduces retry traffic by 73.4\% and p99.5 latency by 36.1\%, improves $q_{\rm del}$ from 0.8263 to 0.8504, and lowers drop rate from 11.62\% to 7.42\%.  
Relative to the fixed 10-hop interval, its retry traffic is within 7.0\% while it uses 57.0\% fewer planned segments and 56.2\% fewer reached checks, lowering p99.5 by 7.5\%.
The \textsc{SAGE}+M negative control keeps the same geographic plan but lets Class-\clM{} failures trigger replay; at $4\times10^{-5}$ this increases retry flit-hops by 86.0\% and worsens $\Psi_{\rm del}$ by 9.7\%.

Figure~\ref{fig:quality_latency_tradeoff} separates delivered quality from mean terminal latency.
\sage improves both over end-to-end, mean-path, and fixed 34-hop recovery.
Fixed 10-hop and \sagep trade higher latency for slightly higher quality; together with \sage, they form the non-dominated set at the plotted means.
\sage has the smallest latency-to-quality ratio at these means: the extra quality from more frequent checkpointing does not offset its delay under the stated objective.

Conversely, \textsc{SAGE}+M is slower and delivers less quality despite retaining the same geographic plan, exposing the cost of allowing bounded Class-\clM{} failures to trigger replay.
Across the six-point paired extension, all ten pairs favor \sage at every BER.
At $3\times10^{-5}$ the paired p99.5 advantage is 410.6 cycles (95\% CI [408.4, 412.9]).  

The complete paired extension and its BER trends appear in Appendix~\ref{app:wrapper:stress}.
Finite retry exhaustion can lower all-outcome tail latency even as recovery load rises.
Because the wrapper does not maintain persistent queues, direct Garnet supplies the headline latency and queue-stability evidence (Sections~\ref{sec:evaluation-direct} and~\ref{sec:evaluation-headroom}).

\subsection{Direct Garnet at the Headline Operating Point}
\label{sec:evaluation-direct}
Table~\ref{tab:sage-direct-synthetic-3e5} compares fixed 34-hop, the path-wide \sagep ablation, and trace-locked \sage under uniform-random traffic and the four-hot-region BER map, using the ordered-VC implementation and campaign contract in Section~\ref{sec:methodology-direct}.
\begin{table}[b]
    \centering
    \caption{
        Direct-Garnet headline and \sagep ablation at $P_{\rm base}=3\times10^{-5}$.
        Entries average per-seed statistics (10 seeds; 15.25M post-warm-up originals per seed).
        Latencies are in cycles.
        Retry and CTRL denote replay DATA and total CTRL flit-hops per original, respectively.
    }    
    \label{tab:sage-direct-synthetic-3e5}
    \small
    \setlength{\tabcolsep}{4.0pt}
    \begin{tabular}{@{}lrrrrrr@{}}
        \toprule
        Policy          & $L_{\rm mean}$    & $L_{99.5}$        & Retry          
                        & CTRL              & Drop              & $q_{\rm del}$     \\
        \midrule
        Fixed 34-hop    & 418.3             & 1269.5            & 17.276         
                        & 14.934            & 11.63\%           & 0.8261            \\
        \textbf{\sage}  & \textbf{301.2}    & \textbf{813.0}    & \textbf{6.418} 
                        & 3.725             & 7.42\%            & 0.8504            \\
        \midrule
        \sagep          & 332.4             & 871.8             & 6.661          
                        & \textbf{3.417}    & \textbf{6.11\%}   & \textbf{0.8592}   \\
        \bottomrule
    \end{tabular}
\end{table}

At $P_{\rm base}=3\times10^{-5}$, \sage lowers mean latency from 418.3 to 301.2 cycles and all-outcome p99.5 from 1269.5 to 813.0 cycles, reductions of 28.0\% and 36.0\% relative to fixed 34-hop.
The mean paired p99.5 advantage is 456.5 cycles (95\% CI $[455.4,457.6]$), and all ten traffic- and seed-paired comparisons favor \sage.
Drop rate falls by 4.21 percentage points, from 11.63\% to 7.42\%, while $q_{\rm del}$ rises from 0.8261 to 0.8504.
Together, the mean-latency and delivered-quality improvements reduce $\Psi_{\rm del}$ by 30.1\%.

\sage plans 26.8\% more segments and executes 19.4\% more endpoint checks than fixed 34-hop, yet retry and control flit-hops fall by 62.9\% and 75.1\%.
Both remain below saturation: fixed 34-hop and \sage have mean maximum directed-link demands of 0.820 and 0.671 and final drains of 1.16\% and 0.71\% of the active-injection interval, respectively.

In the path-wide checkpointing ablation, \sagep achieves slightly higher delivered quality, but region-local \sage lowers $\Psi_{\rm del}$ from 386.9 to 354.2 cycles, an 8.4\% reduction.

\subsection{Direct-Garnet Synthetic High-BER Stress}
\label{sec:evaluation-headroom}

We next test queue stability by increasing BER under the same uniform-random offered load as the headline experiment.
We report five paired direct-Garnet runs at $P_{\rm base}=3\times10^{-4}$, exactly ten times the headline BER, and five paired runs at $10^{-3}$.
At $3\times10^{-4}$, latency metrics aggregate all 15.25M post-warm-up originals; at $10^{-3}$, they use the predeclared 250K cohort after a 4M-original pre-guard.
\begin{table}[t]
    \centering
    \caption{
        Five-seed, traffic- and seed-paired direct-Garnet high-BER stress.
        C/A is terminal-completion/arrival rate and $\Delta B$ is backlog slope; fixed-policy latencies are descriptive nonstationary outcomes.
    }
    \label{tab:sage-direct-synthetic-headroom}
    \begingroup
    \footnotesize
    \setlength{\tabcolsep}{2pt}
    \renewcommand{\arraystretch}{0.96}
    \begin{tabular*}{\columnwidth}{@{\extracolsep{\fill}}lrrrr@{}}
        \toprule
        Policy          & $P_{\rm base}$        & C/A               
                        & $\Delta B$/cyc.       & $L_{\rm mean}/L_{99.5}$   \\
        \midrule
        Fixed 34-hop    & $3\!\times\!10^{-4}$  & 0.558             
                        & +47.71                & 37.8K/179.3K              \\
        \textbf{\sage}  & $3\!\times\!10^{-4}$  & \textbf{1.000}    
                        & \textbf{+0.007}       & \textbf{336.4/759.0}      \\
        Fixed 34-hop    & $10^{-3}$             & 0.486             
                        & +55.51                & 87.7K/214.8K              \\
        \textbf{\sage}  & $10^{-3}$             & \textbf{1.000}    
                        & \textbf{+0.0002}      & \textbf{454.1/1037.0}     \\
        \bottomrule
    \end{tabular*}
    \endgroup
\end{table}

At $P_{\rm base}=3\times10^{-4}$, fixed 34-hop completes only 55.8\% of arrivals and grows backlog by 47.71 originals/cycle, whereas \sage matches arrivals with bounded backlog. Drop rates are nearly matched (26.20\%/25.94\%), yet \sage cuts retry and control traffic by 60.0\% and 74.1\%.

At $P_{\rm base}=10^{-3}$, fixed 34-hop completes 48.6\% of arrivals and grows backlog by 55.51 originals/cycle, while \sage again matches arrivals with bounded backlog. 
At this 26.8\% drop rate, the result establishes extreme overload containment and a $3.3\times$ BER separation between demonstrated BER points at a constant offered load.
Appendix~\ref{app:repro:binning} defines the stationarity diagnostics, and Appendix Table~\ref{tab:app-direct-3e3} gives the extreme-overload point.
%The stable $3\times10^{-5}$ campaign remains the latency headline.

\subsection{Application-Derived Finite-Trace Evaluation}
\label{sec:evaluation-deit}

To test whether the latency benefit extends beyond uniform-random traffic, we evaluate the finite, application-derived DeiT-S communication trace at $r=0.009$.
Results average per-seed statistics over the complete 17.25M-original trace, including the causal controller trajectory and queue history.
At the operating point, $P_{\rm base}=3\times10^{-6}$, ten traffic- and seed-paired comparisons show that \sage lowers mean terminal latency from 284.8 to 237.6 cycles and all-outcome p99.5 from 2657.1 to 714.0 cycles relative to fixed 34-hop.
Mean source-queue time also falls from 25.03 to 4.00 cycles.
With delivered quality nearly matched, $\Psi_{\rm del}$ falls from 287.2 to 239.6 cycles.

At the higher fault-stress point, $P_{\rm base}=10^{-5}$, \sage lowers mean latency from 318.2 to 304.2 cycles and p99.5 from 4006.2 to 3169.2 cycles.
It also cuts the drop rate from 0.05655\% to 0.01713\% and raises $q_{\rm del}$ from 0.956179 to 0.960680.
All ten pairs favor \sage in mean and p99.5 at both lower BERs, and these 40 runs reach zero backlog and protocol quiescence.
Appendix Table~\ref{tab:app-deit-lowrate} reports the complete ten-seed metrics and drain diagnostics; the earlier $r=0.0108$ BER-zero pair remains a separate clean-channel load stress in Appendix~\ref{app:deit-analysis}.

A further ten-seed stress experiment at $P_{\rm base}=2\times10^{-5}$ shows a larger separation under heavier queueing.
Relative to fixed 34-hop, \sage reduces mean terminal latency, p99.5, and $\Psi_{\rm del}$ by 74.2\%, 79.7\%, and 74.6\%, respectively, while raising $q_{\rm del}$ from 0.8966 to 0.9111 and lowering the drop rate from 3.1542\% to 1.3354\%.
All ten pairs favor \sage on these metrics, and all 20 runs reach protocol quiescence (Appendix Table~\ref{tab:app-deit-pb2e5}).

\subsection{Safety Margin and Implementation Cost}
\label{sec:evaluation-safety-cost}

The endpoint qualification uses \sage's ten-seed exposure profile after 6M wrapper warm-up. 
Across 200M CRC32 endpoint-attempt trials, no wrong 128-bit Class-\clH{} payload passes CRC. The worst simultaneous one-sided 95\% per-original-packet bound is $U_{\rm pkt}^{95}=3.18\times10^{-7}<\epsilon_{\rm cat}=10^{-6}$, without assuming independence between attempts or crediting range/non-finite checks.
A same-exposure CRC16 sensitivity records 12 events and remains below the budget with a worst bound of $8.67\times10^{-7}$, but CRC32 retains the larger margin and is the primary design.

The protection expansion is 64 bits per 512-bit semantic flit, or 12.5\%.
Appendix Table~\ref{tab:app-implementation-cost} reports the detailed checker and state costs.
Full pipelining increases combined DC cell area by 34.7\% over the blocking RTL while reducing the Class-\clH{}/Class-\clM{} initiation intervals from 11/5 to 1/1 cycles without changing modeled latency.
Finite checkpoint-store area remains unsynthesized.

\subsection{HBM Portability Pilot}
\label{sec:evaluation-hbm}

In a five-seed random-read pilot at $P_m=10^{-3}$, Class-\clH{}-only recovery preserves normalized semantic goodput $G_{\rm sem}/G_0=0.9959$ with retry factor $\nu=1.0000$, whereas all-class recovery falls to $0.7901$ with $\nu=1.2540$.
This agrees with the NoC negative control: allowing bounded Class-\clM{}/Class-\clL{} outcomes to invoke the expensive path creates substantial recovery traffic.
This wrapper-based pilot supports the portability of semantic recovery to HBM; it does not evaluate a complete two-level controller (Appendix~\ref{app:hbm-portability}).

%% file: figures/fig_tradeoff_q_latency.tex
\begingroup
\begin{tikzpicture}
\begin{axis}[
    width=\linewidth,
    height=0.88\linewidth,
    xmin=0.770, xmax=0.894,
    ymin=235, ymax=520,
    xtick={0.78,0.80,0.82,0.84,0.86,0.88},
    ytick={250,300,350,400,450,500},
    xlabel={Delivered quality $q_{\mbox{\fontsize{8.2}{9}\selectfont del}}$},
    ylabel={Mean terminal latency (cycles)},
    font=\fontsize{9.5}{11}\selectfont,
    tick label style={font=\fontsize{9.5}{11}\selectfont},
    label style={font=\fontsize{9.5}{11}\selectfont},
    xticklabel style={/pgf/number format/fixed,
        /pgf/number format/precision=2,
        /pgf/number format/fixed zerofill},
    scaled ticks=false,
    tick align=outside,
    tick pos=left,
    major tick length=2.5pt,
    axis line style={black!75,line width=0.5pt},
    tick style={black!75,line width=0.5pt},
    clip=false,
    every axis plot/.append style={line width=0.7pt},
]
\tikzset{
    policy label/.style={font=\fontsize{10.5}{12}\selectfont,
        inner sep=1.3pt,fill=white},
    reference label/.style={font=\fontsize{10.5}{12}\selectfont,
        text=black!75,inner sep=1pt,fill=white},
}

% Reference lines satisfy L_mean = Psi_del * q_del.  They do not join policies.
\addplot[black!45,dashed,domain=0.772:0.892,samples=2,no marks] {320*x};
\addplot[black!45,dashed,domain=0.772:0.892,samples=2,no marks] {400*x};
\addplot[black!45,dashed,domain=0.772:0.892,samples=2,no marks] {500*x};
\node[reference label,anchor=south west]
    at (axis cs:0.773,247.36)
    {$\Psi_{\mbox{\fontsize{8.2}{9}\selectfont del}}=320$};
\node[reference label,anchor=south west]
    at (axis cs:0.773,309.2)
    {$\Psi_{\mbox{\fontsize{8.2}{9}\selectfont del}}=400$};
\node[reference label,anchor=south west]
    at (axis cs:0.773,386.5)
    {$\Psi_{\mbox{\fontsize{8.2}{9}\selectfont del}}=500$};

% Lower latency and higher delivered quality are preferred.
\draw[-{Stealth[length=4pt]},black!75,line width=0.7pt]
    (axis cs:0.867,475) -- (axis cs:0.871,420);
\node[font=\fontsize{9.5}{11}\selectfont,text=black!75,
      anchor=south,inner sep=2pt] at (axis cs:0.867,478) {Better};

% Open circles: dominated fixed-policy means.
\addplot[only marks,mark=*,mark size=2.6pt,black,
    mark options={fill=white,draw=black}]
    coordinates {
        (0.8018,479.9) 
        (0.8106,399.9) 
        (0.8263,328.1) 
        (0.7798,280.2)
    };
% Filled diamonds: other non-dominated evaluated means.
\addplot[only marks,mark=diamond*,mark size=3.3pt,black]
    coordinates {
        (0.8626,318.6) 
        (0.8593,290.5)
    };
% Filled circle: SAGE. Marker shapes remain distinct in monochrome.
\addplot[only marks,mark=*,mark size=3.6pt,black,
    mark options={solid,fill=black,draw=black}]
    coordinates {
        (0.8504,270.9)
    };
% Cross: semantic-replay negative control.
%\addplot[only marks,mark=x,mark size=3.0pt,black]
%    coordinates {(0.7798,280.2)};

\node[policy label,anchor=west,xshift=4pt]
    at (axis cs:0.8018,479.9) {End-to-end};
    
\node[policy label,anchor=west,xshift=4pt,yshift=0pt]
    at (axis cs:0.8106,399.9) {Mean-path};
    
\node[policy label,anchor=west,xshift=3pt,yshift=12pt]
    at (axis cs:0.8263,328.1) {Fixed};
\node[policy label,anchor=west,xshift=3pt,yshift=0pt]
    at (axis cs:0.8263,328.1) {34-hop};    
    
\node[policy label,anchor=west,xshift=3pt,yshift=12pt]
    at (axis cs:0.8626,318.6) {Fixed};
\node[policy label,anchor=west,xshift=3pt,yshift=0pt]
    at (axis cs:0.8626,318.6) {10-hop};    
    
\node[policy label,anchor=west,xshift=3pt,yshift=0pt]
    at (axis cs:0.8593,290.5) {SAGE-P};
\node[policy label,anchor=north,xshift=0pt,yshift=-5pt]
    at (axis cs:0.8504,270.9) {\bfseries SAGE};
    
\node[policy label,anchor=west,xshift=3pt,yshift=0pt]
    at (axis cs:0.7798,280.2) {SAGE+M};
\end{axis}
\end{tikzpicture}
\endgroup

%% file: sections/related_work.tex
\section{Related Work}
\label{sec:related-work}

\paragraph{Semantic reliability}
Classical unequal error protection assigns different reliability to message components~\cite{masnick1967linear,albanese1996priority,borade2009uep}; DEC-NoC protects an application-selected MSB prefix and may accept lower-order faults approximately~\cite{chen2018decnoc}.
\sage instead enforces a hard silent-delivery budget for Class-\clH{} and assigns class-specific terminal actions to \clH{}/\clM{}/\clL{}.
Fault-injection studies establish strong bit-position dependence, with sign and exponent faults generally more damaging than mantissa faults~\cite{li2017errorpropagation,mahmoud2020pytorchfi,reagen2018ares,rakin2019bitflip,chen2021ranger}.
Low-precision formats trade representation accuracy for efficiency~\cite{gupta2015limitedprecision,micikevicius2022fp8,rouhani2023microscaling}; \sage operates after quantization and calibrates workload-specific classes rather than changing the format or assuming all mantissa faults are harmless.

\paragraph{Reliable accelerator interconnects}
Prior NoC work covers fault-tolerant routing, link protection, credit-based fabrics, and router redundancy or bypass~\cite{benini2002noc,park2006faulttolerant,agarwal2009garnet,bertozzi2005error,baloch2019defender,rashid2020fault}.
\sage leaves deterministic routing intact but adds ordered VC ownership and exact early-credit accounting; protected NACK and replay traffic consume ordinary network resources and are modeled under contention.

\paragraph{Learning-aware calibration}
Learning curves and scaling laws commonly use exponential-saturation or power-law forms to relate data scale to task performance~\cite{hestness2017predictable,kaplan2020scaling,viering2022learningcurves}.
The residual-to-asymptote term in Eq.~(\ref{eq:sage-learning-law}) is a shifted power law: it is locally exponential-like for small $p_{\mathrm{eff}}$ and has a power-law tail, providing a compact offline calibration across these regimes.
The online \sage controller observes only geographic recovery outcomes.

%% file: sections/conclusion.tex
\section{Limitations and Conclusion}
\label{sec:conclusion}

\begin{comment}
\paragraph{Limits.}
The \clH{}/\clM{}/\clL{} boundary and quality weights are workload- and format-specific.  
The Class-\clH{} bound assumes IID flips under the measured four-hot-region exposure.  
The wrapper is a load-margin estimator rather than a persistent-queue model; direct Garnet executes precomputed causal plan sequences and uses traffic- and seed-paired runs with policy-dependent endpoint outcomes.  
The paired $3\times10^{-4}$ and $10^{-3}$ main-text stresses use one uniform-random load and establish same-BER queue-stability separation only at tested points, not exact thresholds or application headroom.
Fixed-policy latency is nonstationary at both; \sage drops 26.8\% at $10^{-3}$ and 93.31\% at appendix-only $3\times10^{-3}$, so these are stability and overload-containment results rather than useful-service claims.  
The DeiT-S experiment reports full-trace finite-workload results at its own operating and fault-stress points; it does not establish steady state or an absolute saturation threshold.  
Its absolute latencies and BERs are not compared with the uniform-random campaign, and the $P_{\rm base}=0$, $r=0.0108$ run is only a clean-channel load stress.
Ultimately, the NoC evaluation is tightly scoped to isolate its error recovery behavior. 
Consequently, physical checkpoint storage synthesis, temporal or cross-link burst faults, and end-to-end task accuracy under dynamically corrupted payloads remain outside this paper's scope.
\end{comment}

\paragraph{Limits.}
The \clH{}/\clM{}/\clL{} boundary and quality weights are workload- and format-specific, and the Class-H bound assumes IID flips under the measured exposure.
The extreme-BER direct-Garnet experiments demonstrate overload containment rather than useful service.
Correlated burst faults, physical checkpoint-store synthesis, and end-to-end task accuracy under NoC-induced payload corruption remain outside this paper's scope.

\paragraph{Conclusion.}
\sage decouples whether a fault merits replay from where replay restarts.
At the headline operating point, this separation lowers mean latency by 28.0\% and improves delivered semantic quality relative to fixed 34-hop, reducing $\Psi_{\rm del}$---the quality-normalized terminal latency---by 30.1\%.
\sage also sharply reduces retry and control traffic, and all ten paired p99.5 comparisons favor it in a stable regime; under higher-BER stress, it maintains bounded queues whereas the fixed 34-hop baseline becomes nonstationary.
A completed ten-seed DeiT-S experiment extends the evidence beyond uniform-random traffic: \sage lowers mean and p99.5 latency and $\Psi_{\rm del}$ at each evaluated nonzero-BER point.
The CRC32-qualified Class-H path gives a simultaneous 95\% per-original silent-delivery bound of $3.18\times10^{-7}$, below the $10^{-6}$ target, and the HBM pilot supports the portability of selective recovery.
More broadly, semantic replay selection and geographic checkpointing point toward AI data-movement systems that spend recovery bandwidth according to both numerical consequence and physical fault geography.

\section*{Acknowledgments}
\noindent\textbf{Generative AI disclosure.}
Generative AI tools were used to assist with source code development and debugging, manuscript editing and critique, and result consolidation. 
The authors reviewed all AI-assisted materials and take full responsibility for the final content, originality, and technical accuracy of the work.

%% file: appendix/appendix_a.tex
% -----------------------------------------------------------------------------
% Appendix A: Methodology, Calibration, and Reproducibility
% main.tex must issue \appendix before inputting this file.
% Uses array/booktabs/tabularx; acmart supplies the [H] placement used below.
% -----------------------------------------------------------------------------

\clearpage
\section{Methodology, Calibration, and Reproducibility}
\label{app:repro}

This appendix contains the rules of the experiments: campaign populations, the application-derived trace contract, common configuration, implementation scope, manifests, semantic-quality calibration, stationarity criteria, and run-acceptance gates.  Supporting network-level results, including the synthetic extensions and the DeiT-S operating, fault-stress, and clean-channel load-stress results, are collected in Appendix~\ref{app:supplementary}.  Specifically:

\begin{itemize}
    \item \textbf{Campaigns and Populations:} Appendix~\ref{app:repro:campaigns} defines warm-up, pre-guard, measurement populations, seeds, and pairing.
    \item \textbf{Application-Derived Trace:} Appendix~\ref{app:deit-trace} fixes the DeiT-S trace, causal controller trajectory, rate and BER points, reporting population, and interpretation.
    \item \textbf{Configuration:} Appendix~\ref{app:repro:configuration} fixes the network, fault map, and geographic-controller parameters.
    \item \textbf{Implementation:} Appendix~\ref{app:repro:implementation} separates synthesized hardware from modeled state.
    \item \textbf{Manifests and Pairing:} Appendix~\ref{app:repro:manifest} specifies the manifests and pairing checks.
    \item \textbf{Calibration:} Appendix~\ref{app:semantic-detail} defines retained-quality accounting, summarizes calibration, and tests Class-\clH{} boundary promotion.
    \item \textbf{Stationarity:} Appendix~\ref{app:repro:binning} defines queue-stability and saturation diagnostics.
    \item \textbf{Validation Gates:} Appendix~\ref{app:repro:gates} lists the checker, protocol, and trace-validation gates.
\end{itemize}

\subsection{Campaign and Measurement Contracts}
\label{app:repro:campaigns}
\label{app:repro:windows}

Tables~\ref{tab:app-campaigns} and~\ref{tab:app-support-campaigns} summarize campaign preparation and reported populations and make concrete the methodologies in Sections~\ref{sec:methodology-wrapper}--\ref{sec:methodology-supporting}.
Each original enters the declared population once; replay attempts remain attached to it and affect traffic, latency, and terminal outcome without enlarging the denominator.
Reported populations are campaign-specific: wrapper and synthetic NoC statistics exclude their declared warm-up, whereas the DeiT-S operating and fault-stress statistics include the complete finite trace.
Paired direct-Garnet runs share the traffic trace, BER map, executable, and network configuration; adaptive runs consume policy-specific checkpoint sequences precomputed causally over the complete trace.

\begin{table}[t]
    \centering
    \caption{
        NoC campaign and measurement contracts.  
        ``Post-WU'' denotes all originals after simulator warm-up.
    }
    \label{tab:app-campaigns}
    \label{tab:windows}
    \small
    \setlength{\tabcolsep}{3pt}
    \renewcommand{\arraystretch}{1.02}
    \begin{tabular}{@{}
        >{\raggedright\arraybackslash}p{0.24\columnwidth}
        >{\raggedright\arraybackslash}p{0.29\columnwidth}
        >{\raggedright\arraybackslash}p{0.41\columnwidth}@{}}
        \toprule
        Campaign 
        & Preparation and seeds 
        & Reported population and role                                                                      \\
        \midrule
        Wrapper design space
        & Seven policies; 10 seeds; 6M warm-up
        & 250K/policy/BER/seed; four-BER mechanism isolation                                                \\
        Wrapper paired stress
        & Fixed 34-hop and \sage; 10 seeds; 6M warm-up
        & 250K at six BERs; paired p99.5 and $q_{\rm del}$ trend                                            \\
        Direct Garnet, headline
        & Fixed 34-hop/\sage: 10 paired; \sagep: 10; 2M warm-up
        & 15.25M post-WU/run; stable headline/ablation at $3\times10^{-5}$                                  \\
        Direct Garnet, paired $3\times10^{-4}$ stress
        & Fixed 34-hop and \sage; 5 paired seeds; 2M warm-up
        & 15.25M post-WU/run; windowed stability                                                            \\
        Direct Garnet, paired $10^{-3}$ stress
        & Fixed 34-hop and \sage; 5 paired seeds; 2M warm-up; 4M pre-guard
        & Predeclared 250K latency cohort; surrounding run diagnoses stationarity                           \\
        Direct Garnet, DeiT-S operating/fault stress
        & Fixed 34-hop and \sage; 10 paired seeds/BER; causal plan generation over the complete trace; first 6M IDs are 2M warm-up plus 4M pre-guard
        & Full 17.25M-original traces at $3\times10^{-6}$, $10^{-5}$, and $2\times10^{-5}$, $r=0.009$; diagnostics in Appendix~\ref{app:deit-trace} \\
        Direct Garnet, DeiT-S clean load stress
        & Fixed 34-hop and \sage; seed-1 pair; $P_{\rm base}=0$, $r=0.0108$; same simulator run contract
        & 250K latency plus full-trace lifecycle; load-selection diagnostic, not the normal clean baseline  \\
        Direct Garnet, paired $3\times10^{-3}$ extreme stress
        & Fixed 34-hop and \sage; 5 paired seeds; 2M warm-up; 4M pre-guard
        & 15.25M post-WU/run; 5K-window overload-containment diagnostic                                     \\
        \bottomrule
    \end{tabular}
\end{table}

\begin{table}[t]
    \centering
    \caption{
        Qualification and portability campaign contracts.
    }
    \label{tab:app-support-campaigns}
    \small
    \setlength{\tabcolsep}{3pt}
    \renewcommand{\arraystretch}{1.02}
    \begin{tabular}{@{}
        >{\raggedright\arraybackslash}p{0.21\columnwidth}
        >{\raggedright\arraybackslash}p{0.35\columnwidth}
        >{\raggedright\arraybackslash}p{0.38\columnwidth}@{}}
        \toprule
        Campaign 
        & Exposure and seeds 
        & Reported population and role                                          \\
        \midrule
        Endpoint decoder
        & CRC32/CRC16; ten-seed wrapper exposure; five 10M-trial seeds/group
        & Eight BER/CRC groups; simultaneous one-sided 95\% Class-\clH{} bounds \\
        HBM pilot
        & Class-\clH{}-only/all-class; 5 seeds; 40\% clean saturation
        & One million random 64-byte reads/seed; portability                    \\
        \bottomrule
    \end{tabular}
\end{table}

The wrapper has no persistent drain and needs no post-guard.  
Planned segments and reached checks are audited over adjacent late wrapper windows.
The 60 overlapping fixed 34-hop/\sage rows in the four-point matrix and six-point paired sweep agree in every scientific field.
The headline, paired $3\times10^{-4}$, and appendix-only $3\times10^{-3}$ synthetic direct campaigns aggregate all post-WU originals, while the paired $10^{-3}$ campaign reports its predeclared cohort.
The three low-rate DeiT-S campaigns report the complete finite trace, with diagnostic coverage specified in Appendix~\ref{app:deit-trace}; the high-load BER-zero diagnostic reports full-trace lifecycle closure and its predeclared latency cohort, but is not pooled with the $r=0.009$ campaign.
This separates the wrapper load-margin result in Section~\ref{sec:evaluation-wrapper} from the persistent-queue tests in Sections~\ref{sec:evaluation-direct} and~\ref{sec:evaluation-headroom}, and from the finite application-derived operating and stress results in Appendix~\ref{app:deit-analysis}.
Statistics are computed per seed over the named population before seed-level aggregation, following Section~\ref{sec:methodology-statistics}.

\subsection{Application-Derived DeiT-S Trace and Reporting Contract}
\label{app:deit-trace}

The application experiment uses an immutable, rate-shaped communication trace from non-distilled DeiT-S/16 inference on selected ImageNet-1K validation images~\cite{touvron2021deit,russakovsky2015imagenet}.
Twelve physical replicas with six in-flight contexts per replica generate the same packet IDs, routes, and stage-dependent bursts for every paired policy.

\paragraph{Offered-load selection.}
The primary campaign caps aggregate eligibility at 90 originals/cycle, equivalent to $r=0.009$ over the 10,000-router mesh.
A seed-1 scout varied the cap before launching the paired grid.
For the conservative cross-cells used to bracket a common load, the last-injection overhang at $r=0.009$ was 1640 cycles for fixed 34-hop at $3\times10^{-6}$ and 1033 cycles for \sage at $10^{-5}$.
At $r=0.0095$, the corresponding values were 2468 and 17,706 cycles; at $r=0.010$, both exceeded 42,000 cycles.
Thus, 0.009 is the highest tested common rate before abrupt queue accumulation in either cross-cell.
The previous 108-original/cycle cap ($r=0.0108$) is retained only in the BER-zero load-stress diagnostic and is never compared numerically with the lower-rate faulted points.
Rate shaping bounds same-cycle eligibility bursts but does not make the finite trace periodic or establish stationarity.

\paragraph{BER and causal plan generation.}
The operating point is $P_{\rm base}=3\times10^{-6}$; fault stresses at $10^{-5}$ and $2\times10^{-5}$ retain the same offered load.
They are selected independently of the uniform-random BER range because the structured trace has less recovery-traffic headroom.
The $100\times$ hot-region multiplier remains unchanged, giving hot-region BERs of $3\times10^{-4}$, $10^{-3}$, and $2\times10^{-3}$, respectively.
For each BER, the reference controller processes the complete DeiT-S trace in packet-ID order at the matching offered rate and BER.
It emits each packet's plan before accumulating that packet's model-derived Class-\clH{} feedback and updates the source-region table every 40K originals, so feedback affects only later packet IDs.
No uniform-random or higher-BER surrogate is used.

\paragraph{Run and reporting population.}
Each low-rate run captures 17.25M originals: 2M warm-up plus 4M pre-guard supply controller and queue history, followed by the $[6\mathrm{M},6.25\mathrm{M})$ cohort and 11M continued-injection originals.
Causal plan updates continue throughout.
The completed grid contains ten paired seeds at each of the three nonzero BERs, or 60 validated policy runs.
Within each pair, packet IDs, routes, generation cycles, executable, BER map, and network configuration match; checkpoint plans and endpoint outcomes differ.
Reported statistics cover the complete trace, not the separately exported 250K cohort.

At $2\times10^{-5}$, seeds 1--10 vary endpoint faults on one shared traffic trace and one \sage plan.
Host builds share the pinned source and configuration; each pair shares a binary.
Final-quiescence telemetry covers all runs; 5K-cycle windows cover the two lower BERs and seeds 1--3 at $2\times10^{-5}$.

The trace exercises the same queues, credits, checker, NACK, and replay path as the synthetic runs.
Without a verified repetition period, these are finite-workload comparisons, not steady-state estimates or end-to-end inference accuracy measurements.
An absolute saturation threshold requires an offered-load sweep with bounded and persistently growing points.

\subsection{Trace-Locked Controller Configuration}
\label{app:repro:configuration}

Direct Garnet executes exported per-packet checkpoint sequences in trace-locked mode; it neither runs the controller nor feeds realized queue or fault outcomes into later plans.
The offline exporter processes the complete trace causally. It updates $n_s(r)$ every 40K originals, so packet $i$ depends only on earlier model-derived Class-\clH{} feedback.
Tables~\ref{tab:app-noc-config} and~\ref{tab:app-controller-config} expand the common setup in Section~\ref{sec:methodology-common-noc} and the controller/checker realization in Sections~\ref{sec:sage-geographic-controller} and~\ref{sec:implementation-controller}--\ref{sec:implementation-fault-checker}.
Table~\ref{tab:app-noc-config} is common to fixed 34-hop, \sage, and \sagep; Table~\ref{tab:app-controller-config} pins the controller used to generate the two adaptive trajectories.
Trace locking isolates network execution from online controller computation and realized-outcome feedback; it does not freeze one interval table.

\begin{table}[t]
    \centering
    \caption{
        NoC and recovery-protocol configuration.
    }
    \label{tab:app-noc-config}
    \small
    \setlength{\tabcolsep}{3pt}
    \renewcommand{\arraystretch}{1.02}
    \begin{tabular}{@{}
        >{\raggedright\arraybackslash}p{0.35\columnwidth}
        >{\raggedright\arraybackslash}p{0.59\columnwidth}@{}}
        \toprule
        Item & Evaluated setting \\
        \midrule
        Topology and timing             
        & $100\times100$ mesh; 576-bit links; 2-cycle routers; 1-cycle links \\
        DATA / CTRL routing             
        & Deterministic XY / YX \\
        Uniform-random offered / clean-saturation load 
        & 0.0108 / 0.027 packets/router/cycle \\
        DeiT-S primary / clean load-stress rate 
        & 0.009 / 0.0108 packets/router/cycle \\
        Flow control                    
        & 4 VCs/vnet; 8 entries/VC for DATA and CTRL \\
        Checker path                    
        & 16 entries/DATA input, including four checked outputs; one admission/input/cycle when unstalled; 12-cycle surcharge \\
        Retry and arbitration           
        & One original plus at most two replays; at most four consecutive CTRL grants; replay DATA has ordinary priority \\
        Primary handoff                 
        & Ideal local checkpoint handoff; protected NACK and replay traffic remain explicit \\
        \bottomrule
    \end{tabular}
\end{table}

\begin{table}[t]
    \centering
    \caption{
        Source-local causal-controller configuration.
    }
    \label{tab:app-controller-config}
    \small
    \setlength{\tabcolsep}{3pt}
    \renewcommand{\arraystretch}{1.02}
    \begin{tabular}{@{}
        >{\raggedright\arraybackslash}p{0.37\columnwidth}
        >{\raggedright\arraybackslash}p{0.57\columnwidth}@{}}
        \toprule
        Item                        & Evaluated setting \\
        \midrule
        Geography and BER           & 
        $10\times10$ regions, each $10\times10$ routers; hot regions $(2,2)$, $(2,7)$, $(7,2)$, $(7,7)$; receiving-region BER; hot multiplier $100P_{\rm base}$ \\
        Interval state              & 
        $n_0=67$; $n_{\min}=10$; $n_{\max}=199$ hops \\
        Update and score            & 
        40K originals; $\alpha=1/8$; $w_R=10$; $w_D=100$ \\
        Trace-locked generation     & 
        Complete same-trace sequence at each campaign's target BER and offered rate; emit each packet's plan before its feedback; update every 40K originals \\       
        Thresholds                  & 
        $\theta_{\rm high}=10^{-3}$; $\theta_{\rm low}=10^{-4}$ \\
        High-score action           & 
        Halve, bound by $n_{\min}$, and reset $c_{s,r}$ \\
        Low-score action            & 
        Increment $c_{s,r}$; after four such windows add five hops, bound by $n_{\max}$, and reset \\
        Middle band                 & 
        Leave interval and counter unchanged \\
        Attribution and lifetime    & 
        Apportion feedback by regional hop exposure; snapshot endpoints at injection \\
        \bottomrule
    \end{tabular}
\end{table}

The fixed baseline derives its endpoints directly. \sagep applies the smallest source-region interval present when each packet is planned to every route segment, whereas \sage applies the region-local rule in Section~\ref{sec:sage-geographic-controller}; their difference is therefore checkpoint placement, not network configuration.

\subsection{Implementation Realization}
\label{app:repro:implementation}

Table~\ref{tab:app-implementation-cost} expands the cost scope in Sections~\ref{sec:implementation-state} and~\ref{sec:evaluation-safety-cost}.
The Class-\clH{} and Class-\clM{} checker rows are synthesized blocks; the other rows describe bounded state represented in the controller or direct-Garnet model.

\begin{table}[t]
    \centering
    \caption{
        Implementation realization and quantified cost.  
        Checker entries report latency/initiation interval.  
        DC area is in ASAP7 Liberty cell-area units; normalized post-route placed-cell area is in parentheses.
    }
    \label{tab:app-implementation-cost}
    \small
    \setlength{\tabcolsep}{3pt}
    \renewcommand{\arraystretch}{1.02}
    \begin{tabular}{@{}
        >{\raggedright\arraybackslash}p{0.29\columnwidth}
        >{\raggedright\arraybackslash}p{0.65\columnwidth}@{}}
        \toprule
        Component               & Realization and cost / configuration                                          \\
        \midrule
        Class-\clH{} checker    & Synthesizable RTL; 11/1 cycles; 2737.4 ($2949.8~\mu\mathrm{m}^{2}$)           \\
        Class-\clM{} checker    & Synthesizable RTL; 5/1 cycles; 523.3 ($570.4~\mu\mathrm{m}^{2}$)              \\
        Region intervals        & Control state; 100 entries $\times$ 8 bits = 100~B/source                     \\
        Replay snapshot         & Model/controller state; 576 bits (72~B)/active entry, plus metadata           \\
        Checker staging         & 16 entries/DATA input, including four checked outputs, plus one bounded ingress slot/VC  \\
        DATA / CTRL network     & 4 VCs/vnet; 8-entry buffers; CTRL burst limited to four                       \\
        \bottomrule
    \end{tabular}
\end{table}

Finite checkpoint-store area and timing, the ordered ownership ledger, and packet-specific early-credit tokens are modeled but are not included in the standalone checker-area result.
The same ordered DATA/CTRL integration and staging limits apply to fixed 34-hop, \sage, and \sagep.
Thus the reported area is not total router overhead, and the 34.7\% increase in Section~\ref{sec:evaluation-safety-cost} compares fully pipelined with blocking checker RTL only.

\subsection{Manifest and Run Acceptance}
\label{app:repro:manifest}

Every accepted NoC publication run emits a formal manifest, expanding the pairing rules in Section~\ref{sec:methodology-statistics}.
It binds the executable, invocation, inputs, network configuration, and reported packet-ID window before results enter analysis.
A hash mismatch or failed gate excludes the run before aggregation.

\begin{table}[H]
    \centering
    \caption{Fields recorded in the per-run manifest.}
    \label{tab:app-manifest}
    \label{tab:manifest}
    \small
    \setlength{\tabcolsep}{3pt}
    \renewcommand{\arraystretch}{1.02}
    \begin{tabular}{@{}
        >{\raggedright\arraybackslash}p{0.37\columnwidth}
        >{\raggedright\arraybackslash}p{0.57\columnwidth}@{}}
        \toprule
        Recorded fields                             & Verification purpose                                  \\
        \midrule
        Source/prototype revision; executable hash   & Pin source revision and binary                       \\
        Command line; policy; parameters; seed      & Reconstruct invocation and random stream              \\
        BER-map, traffic, checkpoint-plan hashes    & Pin exposure, trace, and policy-specific segmentation \\
        VC, buffer, protocol configuration          & Pin flow-control and recovery capacity                \\
        Packet-ID window; reported population       & Pin measurement and accounting bounds                 \\
    \bottomrule
  \end{tabular}
\end{table}

Across paired policies, traffic, BER-map, executable, and network-configuration hashes must match.  
The packet-ID bounds enforce each campaign's declared reporting population, including the full trace where specified.
Checkpoint-plan identity remains policy-specific: \sage and \sagep consume exported plans, whereas the fixed interval derives its endpoints directly.
Because segmentation changes endpoint sample identities, Section~\ref{sec:methodology-direct} describes fixed 34-hop and \sage as traffic- and seed-paired rather than as receiving identical physical fault masks.

\subsection{Delivered-Quality Accounting and Calibration}
\label{app:semantic-detail}

\paragraph{Fixed retained-quality weights.}
The wrapper and direct-Garnet NoC results use fixed class-level coefficients $\lambda_M=4$ and $\lambda_L=0.25$ across policies and BER points.
For an accepted original $i$,
\begin{equation}
    q_i=\frac{1}{(1+4\rho_{M,i})(1+0.25\rho_{L,i})}.
    \label{eq:app-quality-weights}
\end{equation}
Here $\rho_{M,i}$ and $\rho_{L,i}$ are modeled fractions of the original's 32 BF16-like values with residual damage, not raw BERs or whole-codeword failure probabilities.
The quality coefficients are distinct from the controller weights $w_R,w_D$; Class-\clH{} is governed separately by Eq.~\eqref{eq:sage-catastrophic-constraint}.

\paragraph{Residual accounting.}
Let $p_{i,k}$ be the segment-effective XOR BER defined below for successfully committed segment $k$ of original $i$.
For a value's five Class-\clM{} bits within the 168-bit, single-error-correcting record, the behavioral model uses
\begin{equation}
    \begin{aligned}
        d_M(p)&=1-(1-p)^5-5p(1-p)^{167},\\
        \rho_{M,i}&=1-\prod_k\bigl[1-d_M(p_{i,k})\bigr].
    \end{aligned}
    \label{eq:app-quality-m-residual}
\end{equation}
The last term removes the sole correctable error in the value's five Class-\clM{} bits.
For the seven uncoded Class-\clL{} bits per value, XOR parity is retained across segments:
\begin{equation}
    \begin{aligned}
        p_{L,i}&=\frac{1-\prod_k(1-2p_{i,k})}{2},\\
        \rho_{L,i}&=1-(1-p_{L,i})^7.
    \end{aligned}
    \label{eq:app-quality-l-residual}
\end{equation}
Fractions are clipped to $[0,1]$.
Failed attempts do not accumulate damage: replay restores the preceding checkpoint's quality state.
For \textsc{SAGE}+M, the behavioral checker makes $\rho_{M,i}=0$ on accepted paths.

\paragraph{Segment-effective BER}
Under the independent-flip model, a reached segment $k$ with receiving-region link BERs $P_b(\ell)$ has effective XOR BER
\[
    p_{\oplus,k}
    = \frac{1-\prod_{\ell\in k}\bigl(1-2P_b(\ell)\bigr)}{2}.
\]
The simulator uses this BER to sample endpoint error counts for the 184-bit Class-\clH{}, 168-bit Class-\clM{}, and 224-bit Class-\clL{} fields.

\paragraph{Per-run aggregation.}
For $N$ originals in the declared cohort and accepted subset $\mathcal{A}$, compute
\begin{equation}
    \begin{aligned}
        q_{\rm acc}&=\frac{1}{|\mathcal{A}|}\sum_{i\in\mathcal{A}}q_i,\\
        q_{\rm del}&=\frac{1}{N}\sum_{i\in\mathcal{A}}q_i
                    =(1-P_{\rm drop})q_{\rm acc}.
    \end{aligned}
    \label{eq:app-quality-aggregation}
\end{equation}
Drops score zero; retries are not additional originals.
Set both quality quantities to zero if none is accepted.
Evaluate Eq.~\eqref{eq:app-quality-weights} per original before averaging, not at cohort-mean damage.
In direct Garnet, $q_{\rm del}$ equals the mean exported \texttt{accepted\_quality} in \texttt{packets.csv} over the declared cohort.
Quality and $\Psi_{\rm del}=L_{\rm mean}/q_{\rm del}$ are computed per seed before averaging seeds.

\paragraph{Task-level calibration check.}
Table~\ref{tab:app-qdel-task-check} uses the sign anchor $\lambda_s=4$ and $q_{\rm del}=1/(1+4\rho_s)$, with neither drops nor other injected faults and no per-BER refitting.
This explicitly validates the bounded-residual mapping for this specific architecture, demonstrating the methodology without claiming universal per-bit coefficients.
Table~\ref{tab:app-semantic-evidence} summarizes the fault-injection evidence supporting the semantic class distinction.

\begin{table}[H]
    \centering
    \caption{
        Task-level check of the delivered-quality coordinate on AlexNet/CIFAR-100 at $p=2$.  
        Prediction uses $A_0(2/(1+4\rho_s))$ without per-BER refitting; values are percentage points.
    }
    \label{tab:app-qdel-task-check}
    \small
    \begin{tabular}{@{}crrr@{}}
        \toprule
        $\rho_s$ & Measured & Predicted & Error \\
        \midrule
        0.01 & 62.66 & 62.67 & $+0.01$ \\
        0.03 & 61.60 & 61.58 & $-0.02$ \\
        0.05 & 60.62 & 60.51 & $-0.11$ \\
        0.07 & 59.40 & 59.46 & $+0.06$ \\
        0.09 & 58.41 & 58.42 & $+0.02$ \\
        \bottomrule
    \end{tabular}
\end{table}

\begin{table}[H]
    \centering
    \caption{Representative semantic-calibration evidence.}
    \label{tab:app-semantic-evidence}
    \small
    \begin{tabularx}{\columnwidth}{@{}p{0.23\columnwidth}p{0.31\columnwidth}X@{}}
        \toprule
        Test & Observation & Architectural implication \\
        \midrule
        Exponent MSBs & 2.0\% after activation; 83.0\% with safe restoration & Class-H is filtered by a catastrophic-risk constraint \\
        Lower exponent $e_3$ & $0{\to}1$: 63.38\%; $1{\to}0$: 69.89\% & Class-M is important, directional, and bounded \\
        Mantissa MSB & 69.35\% for $|x|\ge2$ at 3\% BER; near-clean for smaller values & Class-L is magnitude aware but normally non-replaying \\
        \bottomrule
    \end{tabularx}
\end{table}

\paragraph{Class-\clH{} boundary-promotion sensitivity.}
We retain $e_4$--$e_7$ as the minimum Class-\clH{} core and test promoting $e_3$, the sign bit, or both under the same 64-bit budget.
A policy-specific 800M-trial CRC32 campaign observes no silent Class-\clH{} error; the worst per-original upper bound is $U_{\rm pkt}^{95}=4.66\times10^{-7}$.
Under an equal-weight $\lambda=4$ first-order quality sensitivity and the optimistic assumption that every Class-\clH{} size retains the 12-cycle checker, the wrapper estimate of $\Psi_{\rm del}$ is 355.1, 357.3, and 360.6 cycles for zero, one, and two promoted positions, respectively, at $P_{\rm base}=5\times10^{-5}$.
Thus, no tested promotion improves the estimated quality-normalized terminal latency.
Width-dependent checker latency would further penalize promotion, while the corresponding area increase remains a separate implementation cost.

\subsubsection{Illustrative Exponent Recentering}
\label{app:exponent-recentering}

Exponent recentering is an optional transport transformation for tensors whose stored exponents occupy a narrow interval.  For example, the measured AlexNet tensor distribution occupies stored BF16 exponents \(E\in[118,128]\), corresponding to unbiased exponents approximately \([-9,1]\).  
A sender may encode
\[
    E_{\mathrm{tx}} = E + 122,
\]
which maps the occupied interval to \(E_{\mathrm{tx}}\in[240,250]\).
The four most significant transmitted exponent bits are therefore always one over the declared range, eliminating \(0\!\rightarrow\!1\) fault opportunities in precisely the exponent positions that motivate Class-\clH{} protection.  
Because the transformation is one-to-one over the declared exponent interval, it is information preserving: the receiver reconstructs the original BF16 value exactly, with no reduction in precision or represented value set.
The receiver restores \(E=E_{\mathrm{tx}}-122\) before the value is consumed; the recentered representation is never interpreted numerically.

The offset and valid original exponent interval are protected metadata.
The receiver rejects any restoration outside \([118,128]\), including underflow, overflow, or modular wraparound.  Recentering does not replace Class-\clH{} ECC, the end-to-end CRC, or the receiver range checks, and our reported safety and performance results assign it no benefit.  
It is included here only as an example of how a narrow, architecture-measurable exponent distribution can provide an additional representation-level hardening opportunity.

\begin{table}[h]
    \centering
    \caption{
        Illustrative transport-only exponent recentering.
    }
    \label{tab:recenter-example}
    \small
    \begin{tabular}{@{}ll@{}}
        \toprule
        Original stored exponent range  & \(118\)--\(128\)          \\
        Transport offset                & \(+122\)                  \\
        Transmitted exponent range      & \(240\)--\(250\)          \\
        Receiver restoration            & \(E=E_{\mathrm{tx}}-122\) \\
        \bottomrule
    \end{tabular}
\end{table}

\subsection{Stationarity Diagnosis}
\label{app:repro:binning}

This subsection operationalizes the method in Section~\ref{sec:methodology-direct} and the interpretation in Section~\ref{sec:evaluation-headroom}.
A finite trace eventually drains after generation stops, so zero final backlog alone does not establish stationary service; the diagnosis therefore uses complete windows while original traffic remains at full rate.

For a window $w=[t_0,t_1)$ of width $T_w$, define
\begin{equation}
    \begin{aligned}
        B(t)&=N_{\mathrm{gen}}(<t)-N_{\mathrm{term}}(<t),\\
        A_w &= N_{\rm gen}(w)/T_w,\quad C_w=N_{\rm term}(w)/T_w,\\
        \Delta B_w &= [B(t_1)-B(t_0)]/T_w=A_w-C_w.
    \end{aligned}
    \label{eq:backlog}
\end{equation}

The paired $3\times10^{-4}$ diagnosis uses the final five complete full-rate 10K-cycle windows and repeats the test with 5K-cycle bins.  
The paired $10^{-3}$ diagnosis aggregates both policies over cycles 40K--150K and checks the final 50K cycles.
The appendix-only $3\times10^{-3}$ diagnosis uses the final five complete full-rate 5K-cycle windows per seed.
Persistent $C_w<A_w$ with material positive $\Delta B_w$ denotes saturation; matched rates and bounded $B(t)$ denote queue stability.
Because each stress point is evaluated at a single offered load, this classification separates the tested loads into stationary and nonstationary regimes but does not locate an absolute tolerance threshold, which would require an offered-load sweep at fixed BER.
The classification is checked across seeds and windows so opposing slopes cannot cancel in an aggregate mean.
Drop rate is separate because rapid terminal failure can keep a queue bounded while providing poor useful service.
Only queue-stable runs support a steady-state latency interpretation; otherwise quantiles describe the growing queue over the chosen interval.
For the finite DeiT-S trace, generation-boundary backlog, final drain, and late-trace trends diagnose queue accumulation but do not by themselves identify an absolute saturation threshold.  The $r=0.009$ nonzero-BER points and the $r=0.0108$ BER-zero load stress are classified separately and are not combined into a BER trend.

\subsection{Self-Checks and Protocol Gates}
\label{app:repro:gates}

The three gate groups below validate checker timing and backpressure, recovery and ordered credit ownership, and trace identity. 
Failure in any group excludes the run from analysis. 
These gates complement the statistical decoder qualification in Section~\ref{sec:evaluation-safety-cost}.

Publication runs enter analysis only after the full gem5 build and three gate groups complete:
\begin{enumerate}
    \item \textbf{Checker gate:} fixed 11-/5-cycle Class-\clH{}/Class-\clM{} latency, initiation interval one, and output-backpressure stability.
    \item \textbf{Protocol gate:} zero-BER timing; deterministic single- and three-failure NACK/replay/drop tests; one-VC and dense four-VC ordering; and exact early-credit cancellation under forced failure.
    \item \textbf{Trace gate:} traffic, checkpoint plan, executable, BER map, network setup, and measurement-window identities match the manifest.
\end{enumerate}

For the application-derived campaign, the trace gate additionally checks the rate cap, controller-trajectory and input provenance, and exact packet, cycle, source, and destination identities across paired policies.  It also checks packet-lifecycle and ownership/credit-ledger closure.  Appendix~\ref{app:deit-analysis} reports the low-rate operating and fault-stress results and the separate high-load clean-channel diagnostic.

%% file: appendix/appendix_b.tex
\clearpage

\section{Supplementary Evaluation Results}
\label{app:supplementary}
\label{app:synthetic}
\label{app:wrapper}

This appendix collects supporting results and portability pilots that complement Section~\ref{sec:evaluation}. 
It includes the complete Garnet-calibrated wrapper matrices, an extreme-BER direct-Garnet diagnostic, the application-derived DeiT-S operating, fault-stress, and clean-channel load-stress results, and an HBM portability study. 
Specifically:
\begin{itemize}
    \item \textbf{Four-BER Wrapper Matrix:} Appendix~\ref{app:wrapper:matrix} gives the complete four-BER, seven-policy design space.
    \item \textbf{Paired Wrapper Extension:} Appendix~\ref{app:wrapper:stress} gives the six-point fixed 34-hop and \sage comparison.
    \item \textbf{Extreme Direct-Garnet Diagnostic:} Appendix~\ref{app:synthetic:extreme-stress} reports overload containment at $P_{\rm base}=3\times10^{-3}$.
    \item \textbf{Structured-Trace Results:} Appendix~\ref{app:deit-analysis} reports the DeiT-S rate selection, serviceable nonzero-BER points, and high-load clean-channel diagnostic.
    \item \textbf{HBM Portability and Semantic Specialization:} Appendix~\ref{app:hbm-portability} evaluates a Ramulator component pilot to test semantic recovery selection and presents an analytical REACH-style tradeoff for repair spans.
\end{itemize}

The methodology, calibration, and reproducibility are described in Appendix~\ref{app:repro}.

\subsection{Principal Four-BER Matrix}
\label{app:wrapper:matrix}

The matrix contains 280 runs: four principal BERs, seven policies, and ten common seeds.  
Every cell follows the 6M-warm-up plus 250K-measurement contract in Appendix~\ref{app:repro:campaigns}. 

We report retry DATA flit-hops/original as $R_h$ and planned segments/reached checks as P/C.
Latencies are cycles.  
Tables~\ref{tab:app-wrapper-low} and~\ref{tab:app-wrapper-high} are split only for legibility.
\begin{table}[H]
    \centering
    \caption{
        Complete wrapper results at $P_{\rm base}=2\times10^{-5}$ and $3\times10^{-5}$ (ten common seeds).
    }
    \label{tab:app-wrapper-low}
    \small
    \setlength{\tabcolsep}{1.15pt}
    \renewcommand{\arraystretch}{1.00}
    \begin{tabular}{@{}lrrrrrr@{}}
        \toprule
        Policy          & $R_h$             & P/C                   & $L_{\rm mean}$ 
                        & $L_{99.5}$        & $q_{\rm del}$         & Drop              \\
        \midrule
        \multicolumn{7}{@{}l}{\textit{$P_{\rm base}=2\times10^{-5}$}} \\
        End-to-end      & 20.658            & 1.00/1.25             & 396.8 
                        & 2551.8            & 0.8687                & 0.0738            \\
        Mean-path       & 12.902            & 1.49/1.69             & 334.3 
                        & 1414.9            & 0.8812                & 0.0551            \\
        Fixed 34-hop    & 6.131             & 2.45/2.60             & 292.4 
                        & 991.2             & 0.8978                & 0.0316            \\
        Fixed 10-hop    & 1.230             & 7.12/7.20             & 312.8 
                        & 752.1             & 0.9173                & 0.0108            \\
        \sagep          & 1.236             & 4.73/4.81             & 284.2                 
                        & 733.4             & 0.9156                & 0.0108            \\
        \textbf{\sage}  & \textbf{1.377}    & \textbf{2.99/3.08}    & \textbf{264.3} 
                        & \textbf{682.5}    & \textbf{0.9123}       & \textbf{0.0140}   \\
        \textsc{SAGE}+M & 5.247             & 2.99/2.95             & 277.6 
                        & 879.3             & 0.8472                & 0.1478            \\
        \addlinespace[1pt]
        \multicolumn{7}{@{}l}{\textit{$P_{\rm base}=3\times10^{-5}$}}                   \\
        End-to-end      & 29.780            & 1.00/1.38             & 479.9 
                        & 2587.8            & 0.8018                & 0.1508            \\
        Mean-path       & 20.708            & 1.49/1.79             & 399.9 
                        & 1507.6            & 0.8106                & 0.1393            \\
        Fixed 34-hop    & 11.210            & 2.45/2.66             & 328.1 
                        & 1137.3            & 0.8263                & 0.1162            \\
        Fixed 10-hop    & 2.786             & 7.12/7.14             & 318.6 
                        & 785.8             & 0.8626                & 0.0610            \\
        \sagep          & 2.799             & 4.77/4.79             & 290.5 
                        & 773.8             & 0.8593                & 0.0609            \\
        \textbf{\sage}  & \textbf{2.980}    & \textbf{3.06/3.13}    & \textbf{270.9} 
                        & \textbf{726.6}    & \textbf{0.8504}       & \textbf{0.0742}   \\
        \textsc{SAGE}+M & 6.867             & 3.06/2.86             & 280.2 
                        & 944.2             & 0.7798                & 0.2164            \\
        \bottomrule
    \end{tabular}
\end{table}
\begin{table}[H]
    \centering
    \caption{
        Complete wrapper results at $P_{\rm base}=4\times10^{-5}$ and $5\times10^{-5}$ (ten common seeds).
    }
    \label{tab:app-wrapper-high}
    \small
    \setlength{\tabcolsep}{1.15pt}
    \renewcommand{\arraystretch}{1.00}
    \begin{tabular}{@{}lrrrrrr@{}}
        \toprule
        Policy          & $R_h$             & P/C                   & $L_{\rm mean}$ 
                        & $L_{99.5}$        & $q_{\rm del}$         & Drop              \\
        \midrule
        \multicolumn{7}{@{}l}{\textit{$P_{\rm base}=4\times10^{-5}$}}                   \\
        End-to-end      & 34.263            & 1.00/1.44             & 522.6 
                        & 2589.6            & 0.7608                & 0.2013            \\
        Mean-path       & 24.531            & 1.49/1.83             & 433.6             
                        & 1513.5            & 0.7675                & 0.1957            \\
        Fixed 34-hop    & 13.807            & 2.45/2.66             & 345.6 
                        & 1097.7            & 0.7792                & 0.1819            \\
        Fixed 10-hop    & 4.013             & 7.12/6.98             & 318.8 
                        & 792.1             & 0.8125                & 0.1280            \\
        \sagep          & 4.043             & 4.83/4.69             & 291.6                 
                        & 784.8             & 0.8077                & 0.1277            \\
        \textbf{\sage}  & \textbf{4.112}    & \textbf{3.17/3.11}    & \textbf{272.1} 
                        & \textbf{714.9}    & \textbf{0.7982}       & \textbf{0.1429}   \\
        \textsc{SAGE}+M & 7.649             & 3.17/2.92             & 283.6 
                        & 917.9             & 0.7581                & 0.2380            \\
        \addlinespace[1pt]
        \multicolumn{7}{@{}l}{\textit{$P_{\rm base}=5\times10^{-5}$}}                   \\
        End-to-end      & 36.343            & 1.00/1.48             & 542.3 
                        & 2589.6            & 0.7386                & 0.2249            \\
        Mean-path       & 26.217            & 1.49/1.85             & 448.6 
                        & 1486.6            & 0.7448                & 0.2221            \\
        Fixed 34-hop    & 14.910            & 2.45/2.65             & 352.8 
                        & 1010.8            & 0.7547                & 0.2145            \\
        Fixed 10-hop    & 4.707             & 7.12/6.81             & 316.3                 
                        & 772.7             & 0.7786                & 0.1803            \\
        \sagep          & 4.744             & 4.95/4.64             & 290.5 
                        & 757.0             & 0.7725                & 0.1804            \\
        \textbf{\sage}  & \textbf{4.714}    & \textbf{3.31/3.14}    & \textbf{271.6} 
                        & \textbf{679.9}    & \textbf{0.7651}       & \textbf{0.1903}   \\
        \textsc{SAGE}+M & 8.106             & 3.31/3.05             & 287.3 
                        & 911.5             & 0.7502                & 0.2454            \\
        \bottomrule
    \end{tabular}
\end{table}
At the headline $3\times10^{-5}$ point, \sage reduces retry traffic by 73.4\% and p99.5 latency by 36.1\% relative to the fixed 34-hop interval.

Relative to the fixed 10-hop interval, its retry traffic is within 7.0\% while it uses 57.0\% fewer planned segments and 56.2\% fewer reached checks.  
At $4\times10^{-5}$, allowing Class-M failures to invoke replay increases retry flit-hops from 4.112 to 7.649 (86.0\%) and worsens $\Psi_{\rm del}=L_{\rm mean}/q_{\rm del}$ from 340.9 to 374.1 cycles (9.7\%).

\subsection{Paired Fixed 34-Hop/\sage Extension}
\label{app:wrapper:stress}

The paired extension contains 120 runs: two policies, six BERs, and ten common seeds.  
Rows above $5\times10^{-5}$ are detected-failure stress tests outside the CRC-qualified operating envelope.  
Positive $\Delta L_{99.5}$ favors \sage; all ten pairs favor \sage at every listed BER.

\begin{table}[H]
    \centering
    \caption{
        Paired wrapper extension (ten common seeds; 6M warm-up originals).  
        Latencies are all-outcome p99.5 cycles; the final column counts pairs favoring \sage.
    }
    \label{tab:app-wrapper-stress}
    \small
    \setlength{\tabcolsep}{1.35pt}
    \renewcommand{\arraystretch}{1.08}
    \begin{tabular}{@{}ccccc@{}}
        \toprule
        \shortstack{$P_{\rm base}$\\($\times10^{-5}$)}          &
        \shortstack{$L_{99.5}$\\Fixed/\sage}                    &
        \shortstack{$\Delta L_{99.5}$\\\textnormal{(95\% CI)}}  &
        \shortstack{$q_{\rm del}$\\Fixed/\sage}                 &
        Pairs \\
        \midrule
        2  &  991.2/682.5 & \shortstack{308.7\\{[305.4,312.1]}} & 0.8978/0.9123 & 10/10 \\
        3  & 1137.3/726.6 & \shortstack{410.6\\{[408.4,412.9]}} & 0.8263/0.8504 & 10/10 \\
        5  & 1010.8/679.9 & \shortstack{330.8\\{[329.5,332.1]}} & 0.7547/0.7651 & 10/10 \\
        10 &  923.1/660.6 & \shortstack{262.5\\{[261.9,263.2]}} & 0.7056/0.7196 & 10/10 \\
        20 &  938.0/721.0 & \shortstack{217.0\\{[211.7,222.3]}} & 0.6322/0.6808 & 10/10 \\
        50 & 1450.9/748.4 & \shortstack{702.5\\{[699.3,705.7]}} & 0.4476/0.5395 & 10/10 \\
        \bottomrule
    \end{tabular}
\end{table}

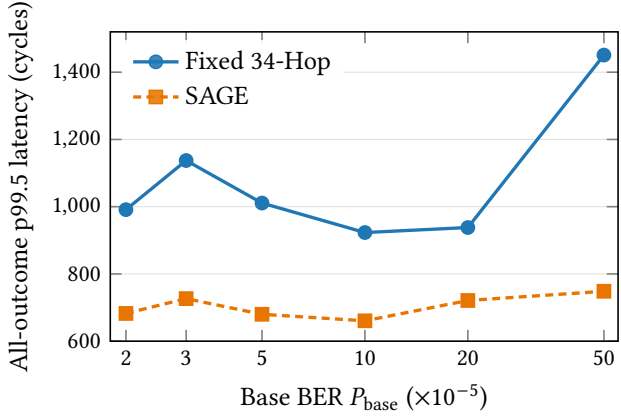
\begin{figure}[H]
    \centering
    \input{figures/wrapper_p995_6m_tikz}    
    \caption{
        All-outcome p99.5 latency in the completed ten-seed, 6M-warm-up wrapper sweep.  
        BER points above $5\times10^{-5}$ are detected-failure stress tests outside the CRC-qualified operating envelope.
    }
    \Description{
        Line plot comparing the fixed 34-hop interval and SAGE all-outcome p99.5 latency across base BER values from 2e-5 to 5e-4.
    }
    \label{fig:sage-wrapper-stress}
\end{figure}

From $5\times10^{-5}$ to $5\times10^{-4}$, a tenfold increase, fixed 34-hop p99.5 rises by 43.5\% while \sage's rises by 10.1\%. 
The corresponding delivered qualities are 0.4476 and 0.5395.  
Table~\ref{tab:app-wrapper-stress} gives the full values, and Figure~\ref{fig:sage-wrapper-stress} plots the latency trend.

The non-monotonicity in Figure~\ref{fig:sage-wrapper-stress} is a consequence of the all-outcome latency metric under finite retry exhaustion. 
For the fixed 34-hop interval, increasing $P_{\rm base}$ from $3\times10^{-5}$ to $5\times10^{-5}$ raises retry flit-hops from 11.210 to 14.910 and drop rate from 11.62\% to 21.45\%, yet lowers all-outcome p99.5 from 1137.3 to 1010.8 cycles.  

As BER rises, more originals exhaust the finite retry budget and terminate before completing the remaining path, turning the all-outcome tail downward.  
Because the wrapper does not maintain persistent queues, the direct-Garnet campaign (Section~\ref{sec:evaluation-headroom}) directly tests high-BER queue stability.

\subsection{Extreme High-BER Direct-Garnet Diagnostic}
\label{app:synthetic:extreme-stress}
\label{app:repro:extreme-stress}

This appendix-only campaign extends Section~\ref{sec:evaluation-headroom} to $P_{\rm base}=3\times10^{-3}$, one hundred times the headline BER, under the same offered load and five paired seeds.
It lies outside the CRC-qualified operating envelope and tests overload containment rather than a useful BER range or an exact saturation threshold.

\begin{table}[H]
    \centering
    \caption{
        Appendix-only extreme direct-Garnet stress at $P_{\rm base}=3\times10^{-3}$ (five traffic- and seed-paired runs).
        C/A, $\Delta B$, and $B_{\rm end}$ use the final five complete 5K-cycle full-rate windows per seed; drop and latency are per-seed means over all 15.25M post-warm-up originals.
        Fixed-policy latencies are descriptive nonstationary outcomes.
    }
    \label{tab:app-direct-3e3}
    \begingroup
    \footnotesize
    \setlength{\tabcolsep}{2.1pt}
    \renewcommand{\arraystretch}{0.98}
    \begin{tabular*}{\columnwidth}{@{\extracolsep{\fill}}lrrrrr@{}}
        \toprule
        Policy          & C/A       & $\Delta B$/cyc.  & $B_{\rm end}$ 
                        & Drop      & $L_{\rm mean}/L_{99.5}$ \\
        \midrule
        Fixed 34-hop    & 0.533     & +50.44           & 7.14M         
                        & 97.87\%   & 66.4K/257.1K            \\
        \sage           & 1.000     & $-0.002$         & 34.6K         
                        & 93.31\%   & 321.4/992.0             \\
        \bottomrule
    \end{tabular*}
    \endgroup
\end{table}

All five fixed 34-hop backlog slopes are positive (50.36--50.53), while the \sage slopes remain near zero ($-0.012$ to $+0.015$).
The reported end-window backlog is 7.14M originals for fixed 34-hop and 34.6K for \sage.
The terminal drop rates, however, are 97.87\% and 93.31\%; thus this point demonstrates bounded protocol and network behavior, not useful service, a latency operating point, application-level headroom, or an exact threshold, consistent with Section~\ref{sec:conclusion}.

\subsection{Structured-Trace Operating and Stress Results}
\label{app:deit-analysis}
\label{app:deit}
\label{app:deit:scope}
\label{app:deit:clean}
\label{app:deit:load-stress}

The immutable DeiT-S communication trace drives the same direct-Garnet queues, credits, checker, protected NACKs, and segment replay as the synthetic campaigns.  
Its traffic structure is application derived, while placement, rate normalization, and the physical mapping remain controlled architectural choices.  
The exact trace, causal controller trajectory, pairing, and reporting contracts appear in Appendix~\ref{app:deit-trace}.  The experiment tests network-level behavior under structured traffic.

\paragraph{Rate selection.}
An initial single-seed sweep selected $r=0.009$ before launching the paired fault campaign.
Table~\ref{tab:app-deit-rate-scout} reports the interval between the final eligible generation and the final first injection.  
The two columns are deliberately conservative cross-cells---the low-BER fixed policy and the high-BER \sage policy---and are used only to locate a common load knee, not to compare policies.

\begin{table}[H]
    \centering
    \caption{
        Seed-1 DeiT-S rate scout.  
        Values are last-injection overhang in cycles.
        The abrupt increase above $r=0.009$ motivates the common primary rate.
    }
    \label{tab:app-deit-rate-scout}
    \small
    \setlength{\tabcolsep}{3.0pt}
    \renewcommand{\arraystretch}{1.04}
    \begin{tabular}{@{}rcc@{}}
        \toprule
        Rate &
        \shortstack{Fixed 34-hop\\$P_{\rm base}=3\times10^{-6}$} &
        \shortstack{\sage\\$P_{\rm base}=10^{-5}$} \\
        \midrule
        0.0090  & 1,640     & 1,033     \\
        0.0095  & 2,468     & 17,706    \\
        0.0100  & 42,883    & 48,271    \\
        \bottomrule
    \end{tabular}
\end{table}

At $r=0.009$, the campaign uses $P_{\rm base}=3\times10^{-6}$ as its application-derived operating point and $10^{-5}$ as a fault stress.  
These BERs are intentionally below the uniform-random campaign because the stage-dependent bursts leave less recovery-traffic headroom; applying the synthetic BER range would primarily measure queue buildup.  
The same four-hot-region map and $100\times$ multiplier are retained, so the local hot BERs are still $3\times10^{-4}$ and $10^{-3}$.

\paragraph{Ten-seed low-rate results.}
\begin{table}[H]
    \centering
    \caption{
        Completed ten-seed full-trace DeiT-S results at $r=0.009$.
        Entries are fixed 34-hop/\sage means of per-seed statistics over 17.25M originals per run.
        Latencies are cycles and quantiles are all-outcome.
    }
    \label{tab:app-deit-lowrate}
    \begingroup
    \fontsize{9}{11}\selectfont
    \setlength{\tabcolsep}{2pt}
    \renewcommand{\arraystretch}{1.03}
    \begin{tabularx}{\columnwidth}{@{}>{\raggedright\arraybackslash}p{0.37\columnwidth}>{\centering\arraybackslash}X>{\centering\arraybackslash}X@{}}
        \toprule
        Metric                      & \shortstack{$P_{\rm base}=3\times10^{-6}$\\Fixed/\sage}
                                    & \shortstack{$P_{\rm base}=10^{-5}$\\Fixed/\sage} \\
        \midrule
        Mean terminal latency       & 284.8 / 237.6
                                    & 318.2 / 304.2             \\
        All-outcome p99.5           & 2657.1 / 714.0
                                    & 4006.2 / 3169.2           \\
        All-outcome p99.9           & 5553.9 / 1901.7
                                    & 7537.6 / 7990.6           \\
        Mean source-queue latency   & 25.03 / 4.00
                                    & 45.11 / 40.57             \\
        Drop rate (total drops)     & $5.80{\times}10^{-7}\%$ (1) / $2.32{\times}10^{-6}\%$ (4)
                                    & 0.05655\% (97,551) / 0.01713\%~(29,555) \\
        $q_{\rm del}$               & 0.991649 / 0.991386
                                    & 0.956179 / 0.960680       \\
        $\Psi_{\rm del}$            & 287.2 / 239.6
                                    & 332.7 / 316.7             \\
        Last-injection overhang     & 1618 / 551
                                    & 1792 / 958                \\
        Final quiescence after last eligibility
                                    & 2046 / 1078
                                    & 2265 / 1522               \\
        \bottomrule
    \end{tabularx}
    \endgroup
\end{table}

At $3\times10^{-6}$, \sage lowers mean, p99.5, p99.9, and source-queue latency by 16.6\%, 73.1\%, 65.8\%, and 84.0\%, respectively, and reduces $\Psi_{\rm del}$ by 16.6\% despite a 0.000263 lower $q_{\rm del}$.
All ten pairs favor \sage on each latency metric; the paired p99.5 advantage is 1943.1 cycles (95\% CI $[1918.5,1967.7]$).
Only one fixed-policy and four \sage originals drop among 172.5M originals per policy.

At $10^{-5}$, \sage lowers mean latency by 4.4\%, p99.5 by 20.9\%, source-queue latency by 10.1\%, drop rate by 69.7\%, and $\Psi_{\rm del}$ by 4.8\%, while raising $q_{\rm del}$ by 0.00450.
All ten pairs favor \sage for mean and p99.5; the paired p99.5 advantage is 837.0 cycles (95\% CI $[773.7,900.3]$).
The p99.9 is instead 453.0 cycles, or 6.0\%, higher under \sage (95\% CI $[271.9,634.1]$), consistently across all ten pairs, so we make no claim at every extreme percentile.

All 40 runs reach zero backlog and protocol quiescence.
The fixed 5K-cycle windows show phase-varying queue buildup and release, while \sage reduces the mean last-injection overhang and time to final quiescence at both BERs.
These diagnostics establish serviceability over the completed finite traces, not an asymptotic saturation threshold.

\paragraph{Extended ten-seed fault stress.}
The additional $P_{\rm base}=2\times10^{-5}$ point retains $r=0.009$ and the same full-trace population.
Table~\ref{tab:app-deit-pb2e5} combines seeds 1--10; all ten pairs favor \sage in mean, p99.5, p99.9, source-queue latency, $q_{\rm del}$, drop rate, and $\Psi_{\rm del}$.

\begin{table}[H]
    \centering
    \caption{
        Extended DeiT-S fault stress at $P_{\rm base}=2\times10^{-5}$ and $r=0.009$: ten paired fault seeds.
        Entries average per-seed full-trace statistics over 17.25M originals per run; latencies are cycles and quantiles are all-outcome.
    }
    \label{tab:app-deit-pb2e5}
    \small
    \setlength{\tabcolsep}{3pt}
    \renewcommand{\arraystretch}{1.02}
    \begin{tabularx}{\columnwidth}{@{}Xrr@{}}
        \toprule
        Metric & Fixed 34-hop & \sage \\
        \midrule
        Mean terminal latency & 4215.1 & 1089.6 \\
        All-outcome p99.5 & 158023.2 & 32122.2 \\
        All-outcome p99.9 & 204845.5 & 140125.4 \\
        Mean source-queue latency & 3674.4 & 762.8 \\
        Drop rate & 3.1542\% & 1.3354\% \\
        $q_{\rm acc}$ & 0.925791 & 0.923424 \\
        $q_{\rm del}$ & 0.896590 & 0.911092 \\
        $\Psi_{\rm del}$ & 4701.3 & 1195.9 \\
        Retry DATA flit-hops/original & 7.878 & 2.250 \\
        CTRL flit-hops/original & 7.200 & 1.782 \\
        Last-injection overhang & 95354.0 & 49870.3 \\
        Final quiescence after last eligibility & 96788.7 & 50568.2 \\
        \bottomrule
    \end{tabularx}
\end{table}

The mean paired advantages are 3125.6 cycles for mean latency (95\% CI $[3081.2,3169.9]$) and 125901.0 cycles for p99.5 (95\% CI $[124275.6,127526.4]$).
The corresponding reductions at the seed means are 74.2\% and 79.7\%; $\Psi_{\rm del}$ falls by 74.6\%.
Although accepted-only quality $q_{\rm acc}$ is slightly lower under \sage in every pair, fewer drops raise $q_{\rm del}$ by 0.014502.
Across 172.5M originals per policy, fixed 34-hop drops 5,440,940 and \sage drops 2,303,593.

All 20 runs terminate with zero backlog and closed protocol ledgers; mean time from final eligibility to quiescence falls from 96788.7 to 50568.2 cycles.
The retained 5K-cycle windows for seeds 1--3 show net backlog growth under fixed 34-hop and net reduction under \sage over cycles 140K--190K in each of those pairs; this windowed observation is not extended to seeds 4--10.
The ten-seed result supports the policy ordering on this finite trace; it does not locate a steady-state saturation threshold.

\paragraph{High-load clean-channel diagnostic.}
\begin{table}[H]
    \centering
    \caption{
        Paired seed-1 DeiT-S clean-channel \emph{load-stress} diagnostic at $P_{\rm base}=0$ and $r=0.0108$.
        Full-trace counts cover 17.25M originals; latency rows use the predeclared 250K cohort $[6\mathrm{M},6.25\mathrm{M})$.
        Latencies are cycles.
    }
    \label{tab:app-deit-load-stress}
    \label{tab:app-deit-clean}
    \small
    \setlength{\tabcolsep}{3pt}
    \renewcommand{\arraystretch}{1.02}
    \begin{tabular}{@{}
        >{\raggedright\arraybackslash}p{0.49\columnwidth}
        rr@{}}
        \toprule
        Metric                                  & Fixed 34-hop  & \sage     \\
        \midrule
        Accepted originals, full trace          & 17.25M        & 17.25M    \\
        Retry DATA / NACK / drop events         & 0 / 0 / 0     & 0 / 0 / 0 \\
        Endpoint checks/original, full trace    & 2.451         & 1.177     \\
        Mean terminal latency, cohort           & 2632.7        & 801.9     \\
        Mean source-queue latency, cohort       & 2244.4        & 521.7     \\
        Delivered-quality loss                  & none          & none      \\
        \bottomrule
    \end{tabular}
\end{table}
The former $r=0.0108$ BER-zero pair is retained as a separate load stress.
Although both policies eventually drain and all originals are delivered, the predeclared cohort is already queue dominated: source waiting accounts for 2244.4 of 2632.7 mean cycles under fixed 34-hop and 521.7 of 801.9 under \sage.  
Because the finite trace has no verified repetition period, this pair is neither a steady-state saturation estimate nor a same-load clean baseline for the $r=0.009$ faulted campaign.  
This diagnostic validates trace and protocol closure at a known queue-dominated load ($r=0.0108$); absolute latencies are evaluated separately in the lower-rate faulted campaign.

\subsection{HBM Portability and Semantic Specialization}
\label{app:hbm-portability}

This study tests whether semantic recovery eligibility remains useful when the contention engine is a memory system rather than a NoC.
Motivated by two-tier and selective HBM ECC~\cite{chen2016rattecc,xie2025hbm}, it combines a Ramulator component pilot with an analytical REACH-style specialization.
The pilot tests recovery selection; the specialization examines protection domains and repair-span costs.
Neither implements a complete two-level HBM controller or a coupled NoC--HBM--compute system.

\subsubsection{Ramulator Component Pilot}
\label{app:hbm-pilot}

\paragraph{Configuration and recovery policies.}
The pilot uses pinned Ramulator~2.0 HBM3~\cite{luo2024ramulator2} with eight channels and one million random 64-byte reads per traffic seed.
Five traffic seeds are evaluated at an original-request load of $0.4r_{\mathrm{sat}}$, where $r_{\mathrm{sat}}$ is the measured clean saturation rate.
Each source line contains $B_{\mathrm{line}}=512$ bits, corresponding to 32 BF16-like values.
The experiment includes an exact $P_m=0$ sanity point and a BER sweep through $P_m=10^{-3}$.
Unlike the NoC study, the fault model has one access exposure per source line and no path-length BER multiplier.

Ramulator supplies memory latency and congestion, while the fixed \clH{}/\clM{}/\clL{} wrapper computes exact-binomial class-failure probabilities, replay amplification, drops, and retained quality.
Recovery-expanded traces account for the additional memory traffic.
The comparison is between Class-\clH{}-only recovery and recovery on any detected \clH{}/\clM{}/\clL{} failure.
Both retain the Class-\clH{} recovery path; the latter also allows bounded Class-\clM{} and Class-\clL{} outcomes to invoke recovery.

For per-attempt recovery-trigger probability $p_{\mathrm{rec}}$ and at most $R$ additional attempts, the independent-attempt wrapper accounting is
\[
    \nu=\sum_{a=0}^{R}p_{\mathrm{rec}}^{a}.
\]
Here $\nu$ includes the initial attempt.
The class-failure and quality terms are analytical; the five traffic seeds vary the random-read workload rather than constituting five independent decoder-safety campaigns.

\paragraph{Goodput and overhead accounting.}
Let $N_{\mathrm{acc}}$ be the number of accepted original lines, $T_{\mathrm{sim}}$ the simulated elapsed time, and $q_{\mathrm{acc,HBM}}$ the wrapper's mean retained-quality score for accepted lines.
Semantic goodput is
\[
    G_{\mathrm{sem}}
    =\frac{B_{\mathrm{line}}N_{\mathrm{acc}}}{T_{\mathrm{sim}}}
     q_{\mathrm{acc,HBM}}.
\]
Dropped lines contribute no useful payload, and replay attempts do not count as additional originals.
We normalize by the clean baseline payload goodput $G_0$.
This throughput-based quantity is distinct from the NoC quality-normalized terminal latency $\Psi_{\mathrm{del}}$.

The native-overhead convention treats storage ECC as part of the baseline memory-reliability layer rather than adding a separate metadata-traffic charge to each policy.
At $P_m=0$, all policies have unit quality, $\nu=1$, zero drops, and $G_{\mathrm{sem}}/G_0=1$ under this convention.
Thus, the pilot isolates semantic recovery selection; it is not a measurement of the incremental parity, packing, or controller cost of the REACH-style design below.
Sidecar and fractional-overhead models charge different metadata traffic and are separate overhead sensitivities.

\paragraph{Result.}
Table~\ref{tab:hbm-portability} reports normalized semantic goodput of 0.9959 for Class-\clH{}-only recovery, with negligible amplification.
All-class recovery increases $\nu$ to 1.2540 and reduces normalized semantic goodput to 0.7901.
The latter spends additional memory traffic recovering bounded residuals.
This is consistent with the NoC semantic-replay negative control: bounded residuals need not automatically invoke the expensive recovery path.

\begin{table}[htbp]
    \centering
    \caption{
        HBM portability pilot at $P_m=10^{-3}$: five random-read traffic seeds, native-overhead accounting.
        Values are seed means; $\nu$ includes the initial access and is rounded to four decimals.
    }
    \label{tab:hbm-portability}
    \begingroup
    \fontsize{9.5}{11}\selectfont
    \setlength{\tabcolsep}{4pt}
    \renewcommand{\arraystretch}{1.06}
    \begin{tabularx}{\columnwidth}{@{}>{\raggedright\arraybackslash}Xrr@{}}
        \toprule
        Recovery trigger                    & $\nu$     & $G_{\mathrm{sem}}/G_0$    \\
        \midrule
        Any \clH{}/\clM{}/\clL{} failure    & 1.2540    & 0.7901                    \\
        Class-\clH{} failure only           & 1.0000    & 0.9959                    \\
        \bottomrule
    \end{tabularx}
    \endgroup
\end{table}

\subsubsection{Analytical REACH-Style Specialization}
\label{app:hbm-reach}

\paragraph{Protection domains.}
The architectural example retains REACH's 32-byte data-transfer unit and inner/outer ECC structure~\cite{xie2025hbm}.
Its inner RS$(36,32)$ code over $\mathrm{GF}(2^8)$ adds four parity bytes to 32 data bytes and corrects up to two unknown byte-symbol errors or four known byte erasures.
The outer code operates over $\mathrm{GF}(2^{16})$ and repairs chunks identified as erasures.
These 32-byte chunks are the units of the span calculation, distinct from the pilot's 64-byte source requests.

Semantic specialization changes which fields invoke the outer path.
Class-\clH{} receives inner correction, outer repair, and range/non-finite validation.
Class-\clM{} receives local correction and a calibrated bounded fallback without default outer repair.
Class-\clL{} is weakly checked or approximately accepted without default outer repair.
The four-bit Class-\clH{} core remains $\{e_4,e_5,e_6,e_7\}$; the repair-span calculation does not require changing the main-text \clH{}/\clM{}/\clL{} boundary.

\paragraph{Outer-span tradeoff.}
An \clH{}-only span uses fewer data bytes, but proportional parity reduction can also reduce multi-chunk erasure capacity.
For a 2048-byte BF16 scalar region, the four Class-\clH{} bit planes occupy
\[
    W_H=2048\,\mathrm{B}\times\frac{4}{16}=512\,\mathrm{B}.
\]
For $W$ data bytes and $P$ outer-parity bytes, the outer RS parameters are $k_{\mathrm{out}}=W/2$ and $n_{\mathrm{out}}=(W+P)/2$.
A known 32-byte chunk erasure consumes 16 outer symbols, giving a full-chunk erasure capacity $C_{\mathrm{era}}=\lfloor P/32\rfloor$.
Table~\ref{tab:hbm-outer-span} compares the full-word span with two \clH{}-only choices for the same original scalar region.

\begin{table}[htbp]
    \centering
    \caption{
        Analytical data-plus-outer-parity spans for the same 2048-byte BF16 scalar region.
        The 32-byte units are burst-equivalent span counts, not measured controller traffic; inner parity, commands, and packing costs are excluded. 
        $C_{\mathrm{era}}$ counts recoverable known 32-byte chunk erasures.
    }
    \label{tab:hbm-outer-span}
    \begingroup
    \fontsize{9.5}{11}\selectfont
    \setlength{\tabcolsep}{3.5pt}
    \renewcommand{\arraystretch}{1.06}
    \begin{tabularx}{\columnwidth}{@{}>{\raggedright\arraybackslash}Xrrrr@{}}
        \toprule
        Domain                      & \shortstack{$W$\\(B)}     & \shortstack{$P$\\(B)}
                                    & \shortstack{32-B\\units}  & $C_{\mathrm{era}}$    \\
        \midrule
        Full word                   & 2048                      & 128 
                                    & 68                        & 4                     \\
        \clH{}, proportional parity & 512                       & 32 
                                    & 17                        & 1                     \\
        \clH{}, retained parity     & 512                       & 128 
                                    & 20                        & 4                     \\
        \bottomrule
    \end{tabularx}
    \endgroup
\end{table}

Scaling parity with the protected data gives a 544-byte \clH{} span, or 17 burst-equivalent units, but reduces the full-chunk erasure budget from four to one.
Retaining 128 parity bytes gives a 640-byte span, or 20 units, while retaining four-chunk erasure capacity.
Thus, reducing repair span is not automatically an equal-reliability comparison.
Keeping a 2048-byte \clH{}-only data span instead would cover four times as many BF16 values and is not the same-region comparison.

The full-word reference here is specifically the 2048-byte data plus 128-byte parity example, not the separate 8/9-rate configuration with 256 parity bytes.
Smaller \clH{}-only spans reduce the bytes involved in an outer recovery; they do not by themselves shorten a fixed-latency decoder pipeline.
Any reduction in repair latency or outer-engine queueing requires a corresponding controller implementation and evaluation.

\subsubsection{Single-Channel Decoder Resource Reference}
\label{app:hbm-logicore}

We separately characterize candidate inner decoders using single-channel AMD LogiCORE RS cores~\cite{amd_rs_decoder} with one 8-bit symbol per cycle over $\mathrm{GF}(2^8)$, with erasure, puncturing, and CCSDS modes disabled.
Table~\ref{tab:hbm-logicore} reports FPGA resources and first-symbol latency $L_1$, measured from the first input symbol to the first corrected output symbol.
These are standalone decoder references, not ASIC area estimates or full-bandwidth HBM implementations.
The $\mathrm{GF}(2^{16})$ outer engine, bus adapters, lane replication, packing logic, and controller buffers are excluded.

\begin{table}[htbp]
    \centering
    \caption{
        Single-channel LogiCORE inner-decoder reference.
        $t$ counts correctable unknown byte-symbol errors; $L_1$ is in core cycles. 
        BRAM is reproduced in the original report's units.
    }
    \label{tab:hbm-logicore}
    \begingroup
    \fontsize{9.5}{11}\selectfont
    \setlength{\tabcolsep}{3pt}
    \renewcommand{\arraystretch}{1.06}
    \begin{tabularx}{\columnwidth}{@{}l>{\raggedright\arraybackslash}Xrrrrr@{}}
        \toprule
        Code & Payload & $t$ & LUTs & FFs & BRAM & $L_1$ \\
        \midrule
        RS$(36,32)$ & Full & 2 & 458 & 401 & 1.0 & 77 \\
        RS$(20,16)$ & 16-byte subset & 2 & 485 & 397 & 1.0 & 61 \\
        RS$(10,8)$  & \clH{} only & 1 & 400 & 317 & 0.5 & 36 \\
        \bottomrule
    \end{tabularx}
    \endgroup
\end{table}

Evaluating a 16-byte subset payload demonstrates the hardware scaling behavior of a reduced-width semantic configuration without redefining the core \clH{}/\clM{}/\clL{} bounds evaluated in the NoC study.
A combined \clH{}/\clM{} code also does not provide separate \clH{}/\clM{} failure flags by itself.

The two $t=2$ cores use similar resources: shortening the protected payload from 32 to 16 bytes does not demonstrate an area reduction.
It concentrates the same four parity bytes on fewer data symbols and reduces $L_1$ in this single-channel configuration.
RS$(10,8)$ uses fewer resources but corrects only one unknown byte-symbol error, so it is not an equal-strength replacement.
These results motivate examining protection density and recovery traffic rather than assuming that semantic specialization automatically reduces decoder area.

\paragraph{Implementation requirements and scope.}
Strict \clH{}-only escalation requires independently interpretable Class-\clH{} validation.
A shared \clH{}/\clM{} code with one uncorrectable flag cannot establish that \clH{} is intact when only \clM{} is to be approximately accepted.
Separate \clH{}/\clM{} inner codes or an independent \clH{} validation mechanism are therefore needed.
Field packing, swizzle/deswizzle logic, outer-parity updates, and separate failure reporting remain additional integration costs.

The random-read pilot does not implement these mechanisms, qualify the proposed RS hierarchy against silent Class-\clH{} errors, or evaluate end-to-end task accuracy.
Persistent-state faults may require reload from a clean copy, scrubbing, or checkpoint recovery rather than an ordinary repeated access.
This evidence confirms the portability of semantic recovery eligibility to memory-system contention models, establishing a foundation for future REACH-style HBM controller implementations.

%% file: figures/wrapper_p995_6m_tikz.tex
% Figure 2 plot body (PGFPlots/TikZ).
% Required in the document preamble:
%   \usepackage{pgfplots}
%   \pgfplotsset{compat=1.18}
%
% Font sizes are explicit so that the figure remains legible in an
% ASPLOS one-column layout. Adjust these three commands if needed.

\begin{comment}
\providecommand{\FigTwoTickFont}{\fontsize{8.5}{10}\selectfont}
\providecommand{\FigTwoLabelFont}{\fontsize{9}{10.5}\selectfont}
\providecommand{\FigTwoLegendFont}{\fontsize{8.5}{10}\selectfont}
\end{comment}

\providecommand{\FigTwoTickFont}{\fontsize{9}{10.5}\selectfont}
\providecommand{\FigTwoLabelFont}{\fontsize{10}{11.5}\selectfont}
\providecommand{\FigTwoLegendFont}{\fontsize{10}{11.5}\selectfont}

\definecolor{SAGEFixedBlue}{HTML}{1F77B4}
\definecolor{SAGEOrange}{HTML}{E87500}

\begin{tikzpicture}
    \begin{axis}[
        width=0.98\linewidth,
        height=0.67\linewidth,
        xmode=log,
        log basis x=10,
        xmin=1.8e-5,
        xmax=5.5e-4,
        ymin=600,
        ymax=1520,
        xtick={2e-5,3e-5,5e-5,1e-4,2e-4,5e-4},
        xticklabels={2,3,5,10,20,50},
        ytick={600,800,1000,1200,1400},
        xlabel={Base BER $P_{\rm base}$ ($\times 10^{-5}$)},
        ylabel={All-outcome p99.5 latency (cycles)},
        tick label style={font=\FigTwoTickFont},
        label style={font=\FigTwoLabelFont},
        legend style={
            font=\FigTwoLegendFont,
            at={(0.025,0.975)},
            anchor=north west,
            draw=none,
            fill=white,
            fill opacity=0.88,
            text opacity=1,
            cells={anchor=west},
            row sep=1pt
        },
        axis line style={line width=0.6pt},
        tick style={line width=0.6pt},
        ymajorgrids=true,
        grid style={gray!25,line width=0.35pt},
        enlarge x limits=false,
        clip marker paths=true,
    ]

    \addplot+[
        color=SAGEFixedBlue,
        solid,
        line width=1.25pt,
        mark=*,
        mark size=2.5pt,
        mark options={solid,line width=0.55pt}
    ] coordinates {
        (2e-5,991.2)
        (3e-5,1137.3)
        (5e-5,1010.8)
        (1e-4,923.1)
        (2e-4,938.0)
        (5e-4,1450.9)
    };
    \addlegendentry{Fixed 34-Hop}

    \addplot+[
        color=SAGEOrange,
        densely dashed,
        line width=1.25pt,
        mark=square*,
        mark size=2.5pt,
        mark options={solid,line width=0.55pt}
    ] coordinates {
        (2e-5,682.5)
        (3e-5,726.6)
        (5e-5,679.9)
        (1e-4,660.6)
        (2e-4,721.0)
        (5e-4,748.4)
    };
    \addlegendentry{SAGE}

  \end{axis}
\end{tikzpicture}